\documentclass[11pt]{article}
\usepackage[utf8]{inputenc}
\usepackage[T1]{fontenc}
\usepackage[english]{babel}

\usepackage{graphicx} 
\usepackage[labelfont={sf,bf}]{caption}
\usepackage{subfig}
\usepackage[margin=0.8in]{geometry} 
\usepackage{amsmath,amssymb,bm} 
\usepackage[dvipsnames,svgnames,x11names]{xcolor} 

\definecolor{Rstring}{HTML}{008000}
\definecolor{Rcomments}{HTML}{4D4D4D}
\usepackage{listings}
\lstdefinelanguage{myRlanguage}{
  morekeywords={library, FALSE, TRUE}, 
  numberstyle = \color{blue}, 
  alsoletter={.}, 
  morestring=[b][\color{Rstring}]", 
  morecomment = [l][\color{Rstring}]{\#}
}
\lstdefinestyle{myRstyle}{
  language = {myRlanguage}, 
  keywordstyle = \bfseries\color{blue}, 
  showstringspaces=false, 
  basicstyle = \footnotesize\ttfamily, 
  moredelim=**[is][\color{red}]{@}{@}
}

\usepackage{enumitem}

\usepackage{booktabs}
\usepackage[tight-spacing=true]{siunitx}
\usepackage{multirow}

\usepackage{todonotes}

\usepackage[ruled, vlined]{algorithm2e}

\usepackage{xpatch}
\usepackage[autostyle,italian=guillemets]{csquotes}
\usepackage[backend=biber,style=authoryear-comp, sortcites, sorting=ynt, mincitenames=1, uniquelist=false]{biblatex}
\usepackage{guit}
\usepackage{xurl}
\PassOptionsToPackage{unicode}{hyperref}
\PassOptionsToPackage{hyphens}{url}
\usepackage{bookmark}
\hypersetup{
  pdftitle={A non-separable first-order spatio-temporal intensity for events on linear networks: an application to ambulance interventions},
  pdfauthor={Andrea Gilardi, Riccardo Borgoni and Jorge Mateu},
  colorlinks=true,
  linkcolor={RoyalPurple},
  filecolor={Maroon},
  citecolor={Blue},
  urlcolor={Blue}
}
\newcommand{\footremember}[2]{%
    \footnote{#2}
    \newcounter{#1}
    \setcounter{#1}{\value{footnote}}%
}
\newcommand{\footrecall}[1]{%
    \footnotemark[\value{#1}]%
}

\title{A non-separable first-order spatio-temporal intensity for events on linear networks: an application to ambulance interventions}
\author{Andrea Gilardi\footremember{corr}{Corresponding author. Email: andrea.gilardi@unimib.it}\footremember{alley}{University of Milano - Bicocca} \and Riccardo Borgoni\footrecall{alley} \and Jorge Mateu\footremember{trailer}{Universitat Jaume I}}
\date{Last compiled on \today}

\begin{document}

\maketitle
\begin{refsection}
\begin{abstract}
The algorithms used for the optimal management of an ambulance fleet require an accurate description of the spatio-temporal evolution of the emergency events. In the last years, several authors have proposed sophisticated statistical approaches to forecast ambulance dispatches, typically modelling the data as a point pattern occurring on a planar region. Nevertheless, ambulance interventions can be more appropriately modelled as a realisation of a point process occurring on a linear network. The constrained spatial domain raises specific challenges and unique methodological problems that cannot be ignored when developing a proper statistical approach. Hence, this paper proposes a spatio-temporal model to analyse ambulance dispatches focusing on the interventions that occurred in the road network of Milan (Italy) from 2015 to 2017. We adopt a non-separable first-order intensity function with spatial and temporal terms. The temporal dimension is estimated semi-parametrically using a Poisson regression model, while the spatial dimension is estimated non-parametrically using a network kernel function. A set of weights is included in the spatial term to capture space-time interactions, inducing non-separability in the intensity function. A series of tests show that our approach successfully models the ambulance interventions and captures the space-time patterns more accurately than planar or separable point process models. 
\end{abstract}


\section{Introduction}
\label{sec:intro}

The proper management of an ambulance fleet is of vital importance for the timely assistance of medical emergencies, particularly when, as the latest COVID-19 pandemic has demonstrated, healthcare operations are stressed by long-standing critical events, such as epidemics or natural and man-made disasters. Relevant efforts are devoted by local agencies to allocate limited human and instrumental resources while managing increasing demand for services, guaranteeing high levels of geographical coverage and a constant improvement of key performance metrics such as rapid responses to potentially life-threatening emergencies \parencite{vile2012predicting}.

Policymakers require qualitative and quantitative approaches and evidence-based studies to tackle these challenging issues. In fact, the management of an Emergency Medical System (EMS) is an extremely difficult task considering the complex spatial and temporal dynamics that govern ambulance interventions, especially for large and highly populated metropolitan areas. Advanced operational research algorithms have been developed in the past years to manage the fleet size and locate the dispatch centres \parencite{blackwell2002response, henderson2011operations}. However, these algorithms depend upon ad-hoc inputs regarding the distribution of emergency events, and the adoption of inaccurate predictions can lead to poor deployment decisions, high response times, and, in general, low performances. Therefore, in the past years, several authors (see, e.g. \textcite{zhou2015predicting, zhou2015spatio, bayisa2020large}) proposed complex spatio-temporal models to carefully forecast the interventions. 

Typically, in the aforementioned papers, the ambulance dispatches were modelled as point processes occurring on a planar surface (e.g. a polygon delimiting a city). Nevertheless, we believe that the emergency interventions can be more appropriately considered as a realisation of a point process occurring on a linear network, i.e. a graph object whose nodes and edges are embedded in a space \parencite{barthelemy2011spatial, baddeley2021analysing}. Street networks represent a particular case of linear networks where nodes and edges correspond to road junctions and street segments, respectively. 

The analysis of spatial data occurring on a linear network raises geometrical, computational, and statistical complexities \parencite{okabe2012spatial}. First, ignoring the network constraint may lead to spurious results and false positive detections \parencite{yamada2004comparison, lu2007false}. Second, the re-adaptation of the classical planar techniques (such as the K-function or the kernel density estimator) presents unique methodological problems due to the non-homogeneous nature of the spatial domain\footnote{A street network is not a homogeneous spatial domain since each edge (i.e. each road segment) is surrounded by different configurations of the neighbouring streets.}. Third, the length of the spatial network and volume of the data typically create additional computational problems that require ad-hoc solutions \parencite{rakshit2019fast, rakshit2019efficient}. We refer to \textcite{baddeley2021analysing} and the references therein for more details.

The goal of this paper is to analyse all ambulance interventions that occurred in Milan (Italy) from 2015 to 2017 using a spatio-temporal point pattern model developed at the road network level. Starting from the assumption that the emergency events can be modelled by a non-homogeneous Poisson Process, we propose a non-separable structure for the first-order intensity function with spatial and temporal terms. The temporal component is modelled semi-parametrically using a Poisson regression with deterministic covariates, while the spatial dimension is modelled using a non-parametric kernel estimator. The non-separability of the intensity function is induced by a set of weights that are included in the spatial component to capture space-time interactions. To the best of our knowledge, this paper represents the first attempt to model EMS data on an extensive road network via a non-separable intensity function. 

The rest of the paper is organised as follows. Section~\ref{sec:data} examines the ambulance interventions data and presents the procedures used to build the computational representation of the street network. We introduce the spatio-temporal framework and the first-order non-separable intensity function in Section~\ref{sec:methods}, providing an overview of the spatial and temporal statistical models. The main results are presented in Section~\ref{sec:results}, while, in Section~\ref{sec:validation}, we validate the performances of the proposed methodology. Section~\ref{sec:compare} compares the suggested approach with alternative specifications that include planar or separable approaches, discusses the scalability of our model to large linear networks and exemplifies a real-world application. Finally, Section~\ref{sec:conclusions} concludes the article summarising the most important findings and the main contributions. 

\section{Data: Ambulance interventions}
\label{sec:data}

\begin{figure}[tb] 
  \centering 
  \subfloat[Year: 2015\label{fig:spatial-distribution-1}]{
    \includegraphics[width=0.4\linewidth]{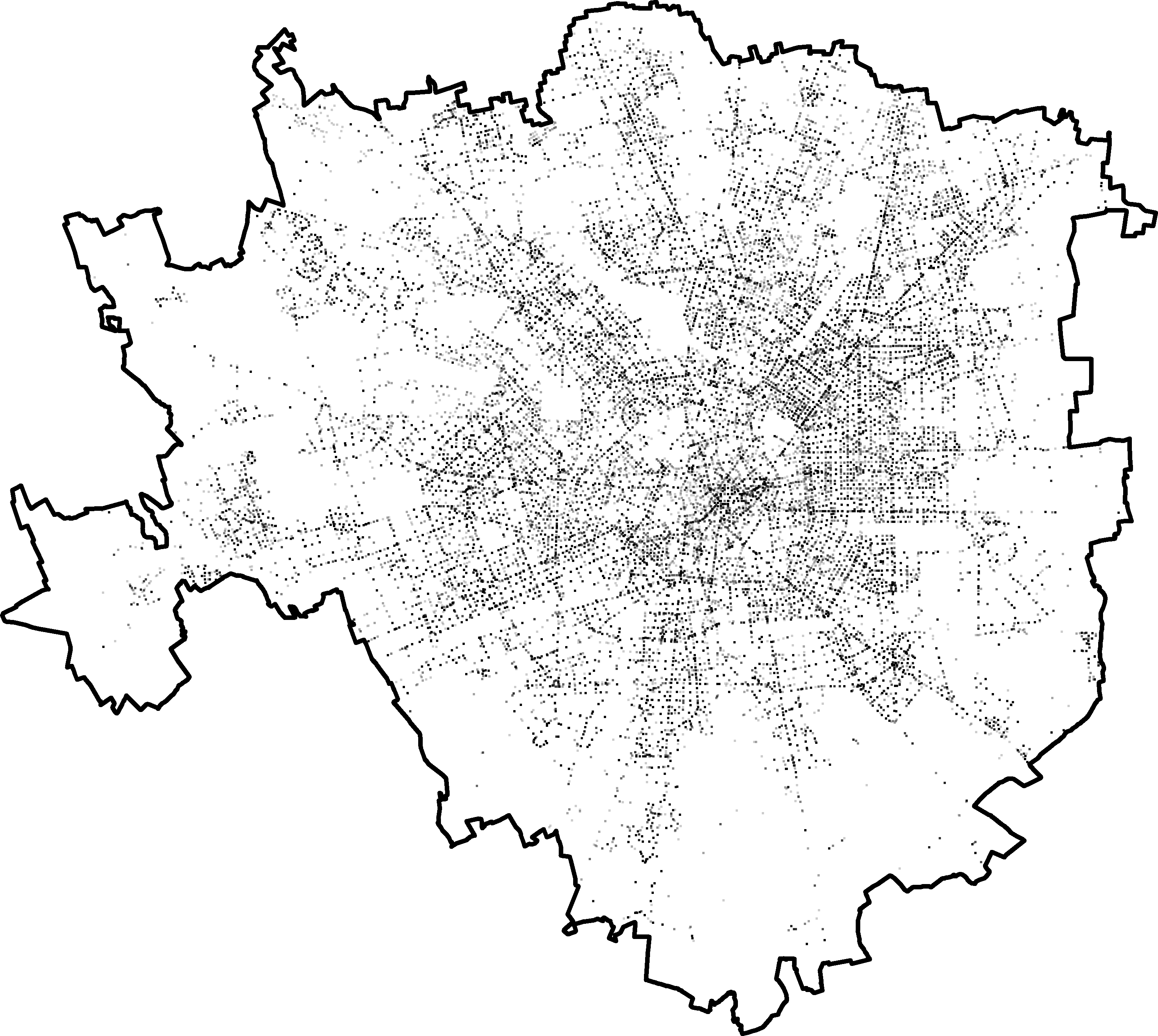}
  }
  \hspace*{1.5cm}
  \subfloat[Year: 2016\label{fig:spatial-distribution-2}]{
    \includegraphics[width=0.4\linewidth]{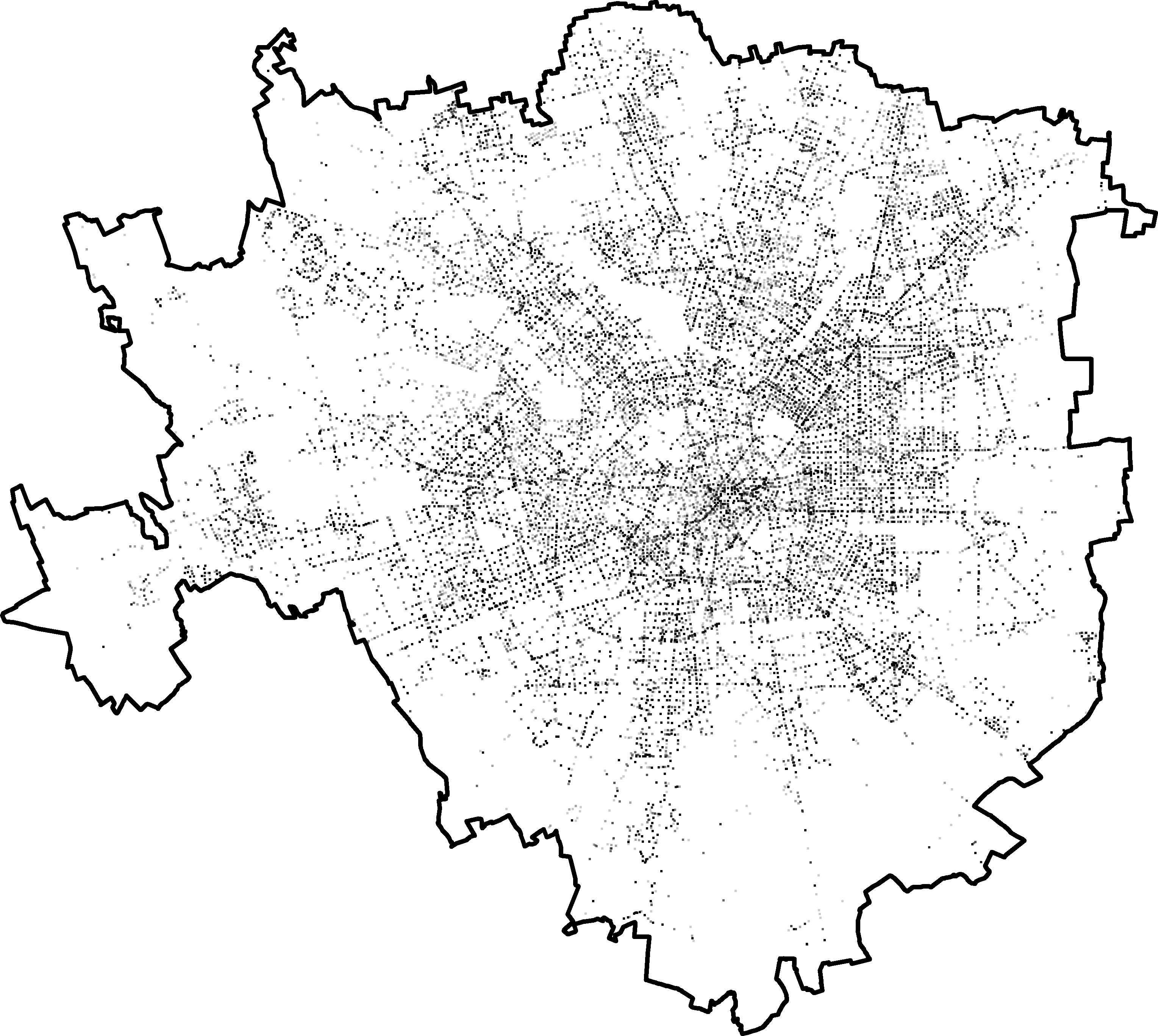} 
  }\\
  \subfloat[Year: 2017\label{fig:spatial-distribution-3}]{
    \includegraphics[width=0.4\linewidth]{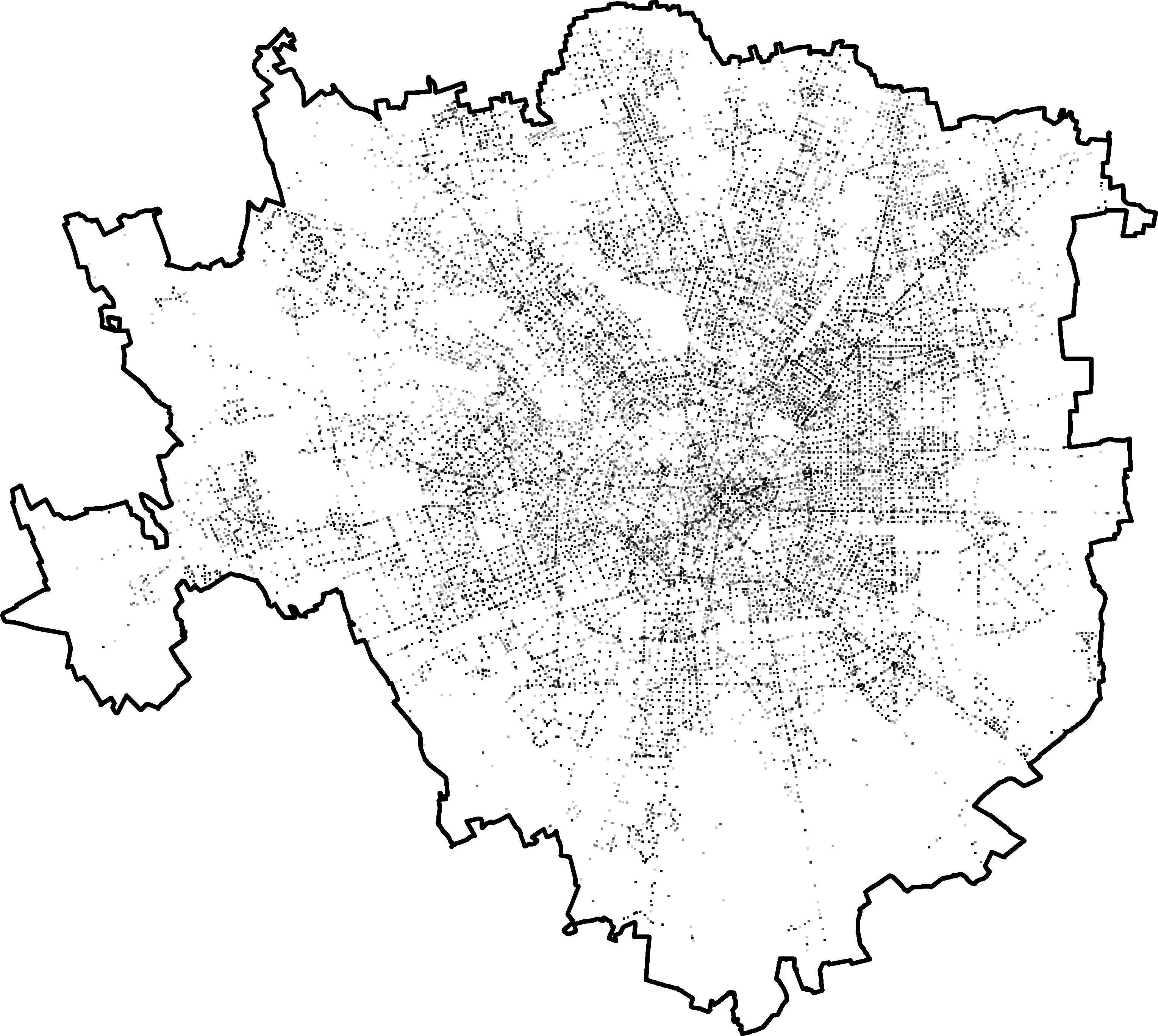}
  } 
  \hspace*{1.5cm}
  \subfloat[
    Milan's road network\label{fig:spatial-distribution-4}
  ]{
    \includegraphics[width=0.4\linewidth]{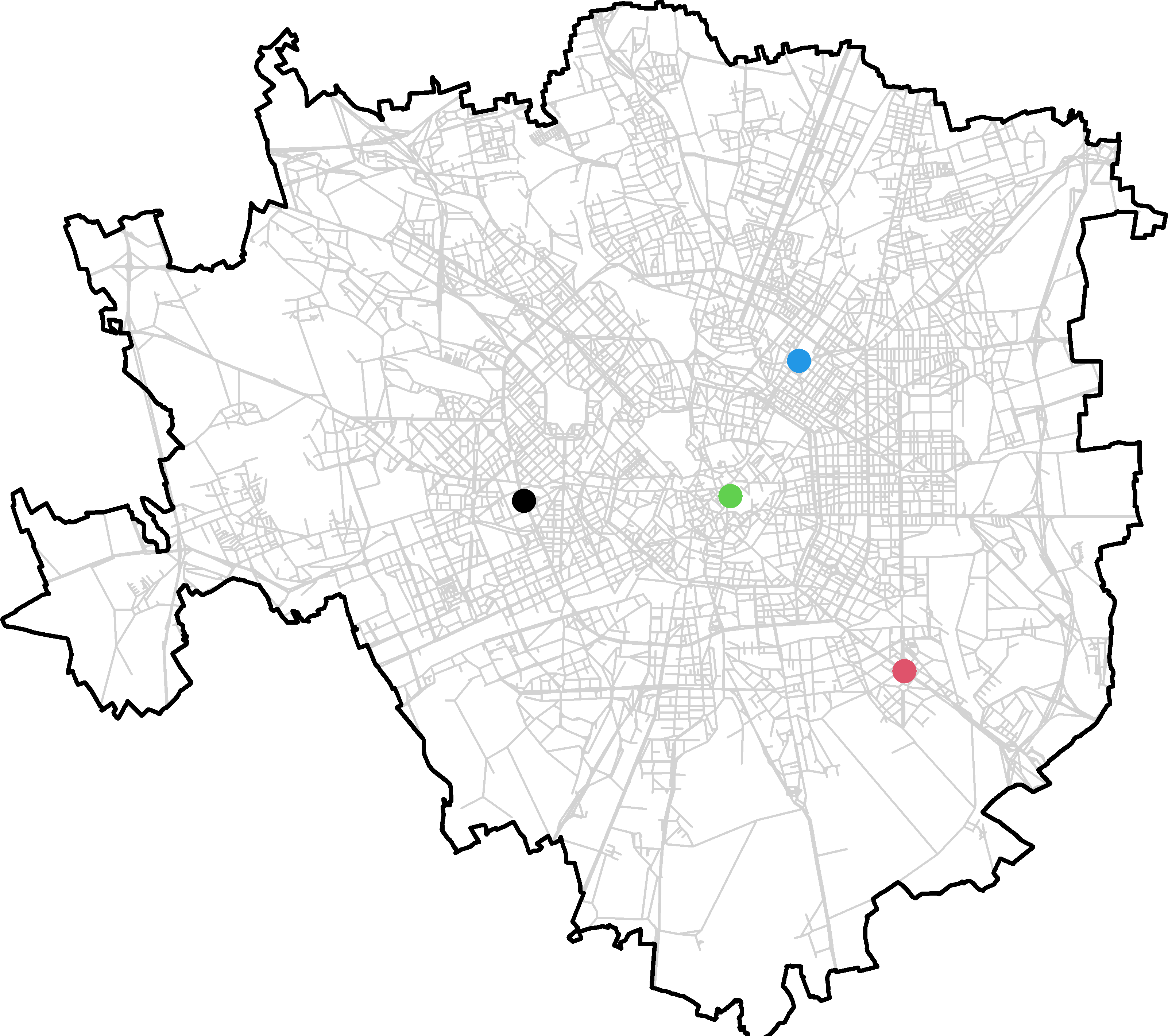} 
  }
  \caption{(a) to (c): locations of ambulance interventions that occurred in Milan from 2015-01-01 to 2017-12-31. Each map represents one year; (d): Milan's road network. The blue dot denotes \emph{Stazione Centrale} (i.e. the Central Station), the black dot denotes an important retirement house, the green dot denotes the Duomo of Milan, and the red dot denotes an important square.}
  \label{fig:spatial-distribution}
\end{figure}

In this Section, we describe the characteristics of the data, its peculiarities, and the procedures used to transform the raw data into a computational structure suitable for fitting the proposed model, which is detailed in Section \ref{sec:methods}.

The dataset was provided by the official regional EMS and included all ambulance dispatches in the city of Milan from 2015-01-01 to 2017-12-31. Milan is the largest city in Italy after Rome, with a total population of 1,386,235 in 2021 and an area of about 183 square kilometres. It represents one of the most important Italian metropolitan areas, where hundreds of thousands of people pass through every day. For this reason, the management of the ambulances in Milan requires ad-hoc modelling and planning. The dataset includes four fields recording the day, the hour, and the GPS coordinates of the ambulance interventions (stored using an official Italian coordinate reference system named Gauss-Boaga projection). These columns provided the necessary information to estimate the spatio-temporal model introduced below. 

Approximately 50,000 observations representing errors, outliers, and spurious or duplicated ambulance dispatches were excluded from the raw data. More precisely: 
\begin{enumerate}[noitemsep, nolistsep]
	\item we did not consider 10,000 outlier interventions that occurred during EXPO 2015, the World Expo hosted in Milan from May 1 to October 31 dedicated to food and life themes. The event attracted about 2.15 million visitors from all around the world and the emergency interventions that occurred in the area where EXPO took place were handled in a different way than usual;
	\item we ignored 11,500 erroneous calls (i.e. situations where someone requested an ambulance but an error was recorded in the EMS database);
	\item we removed 30,000 records linked to multiple ambulance dispatches. These situations typically occur in particular circumstances, like serious or life-threatening emergencies (e.g. heart failures or severe car crashes). In these cases, duplicated observations are recorded for the same event, but we retained only the first ambulance dispatch;
	\item we filtered out 7,000 observations linked to ambulance reroutings (i.e. an ambulance going to location A gets redirected to another, typically more pressing, emergency at location B), and we considered only those records that correspond to the actual interventions.   
\end{enumerate}

Similar preprocessing steps were also implemented in \textcite{matteson2011forecasting, zhou2015predicting}. The remaining sample included 494614 interventions, 163075 occurred in 2015, 164871 in 2016, and 166668 in 2017.

The spatial distribution of the EMS events is depicted in Figures \ref{fig:spatial-distribution-1}-\ref{fig:spatial-distribution-3}. The spatial patterns look stable among the three years and the interventions clearly mirror the skeleton of a road network (see Figure \ref{fig:spatial-distribution-4}), highlighting the city ring road and some of the most important arterial thoroughfares. Most of the white areas represent non-urban places, mainly located in the southern and western parts of the municipality. Given the spatial distribution of the emergency interventions, we believe that a network-approach is more appropriate than a planar approach since it takes into account the nature of the data and the particular constrains of their geo-locations.
 
\begin{figure}[tb]
  \centering
  \includegraphics[width=0.85\linewidth]{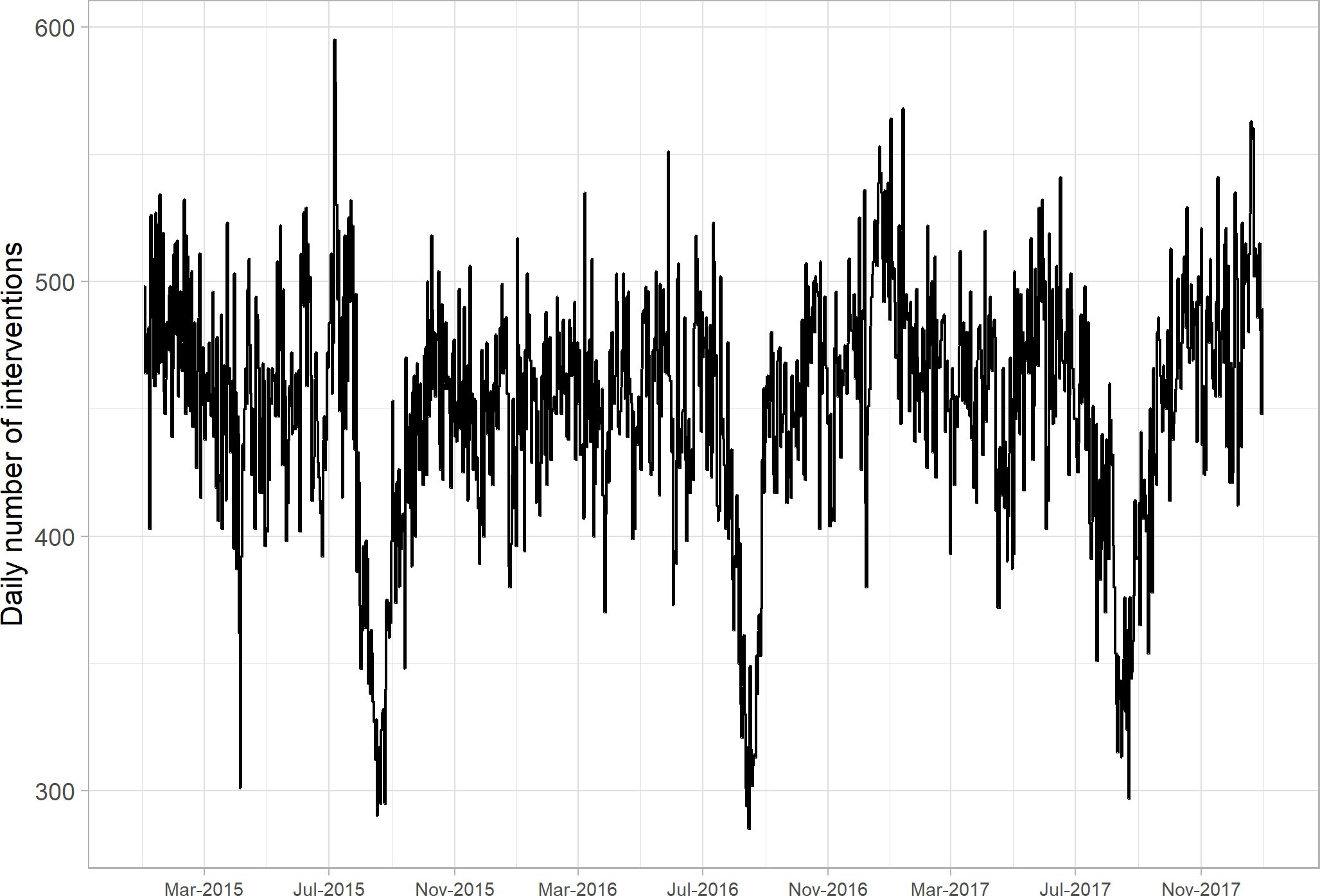}
  \caption{Daily number of ambulance interventions that occurred in Milan from 2015-01-01 to 2017-12-31.}
  \label{fig:daily-events}
\end{figure}

We also explored the temporal dimension of the data, examining the daily and weekly dynamics that govern the total number of emergency interventions. Figure \ref{fig:daily-events} shows the daily number of ambulance dispatches. The data exhibit a clear trend, with a global minimum registered in August, the typical period for summer holidays in Italy. Other local peaks and minima could be linked with the most important religious holidays (such as Christmas or Easter), national celebrations (New year’s eve) or other occasional events (such as the heatwave in July 2015 or the ice storms in January 2017). The three years are characterised by similar temporal patterns.

\begin{figure}
  \centering
  \includegraphics[width=0.85\linewidth]{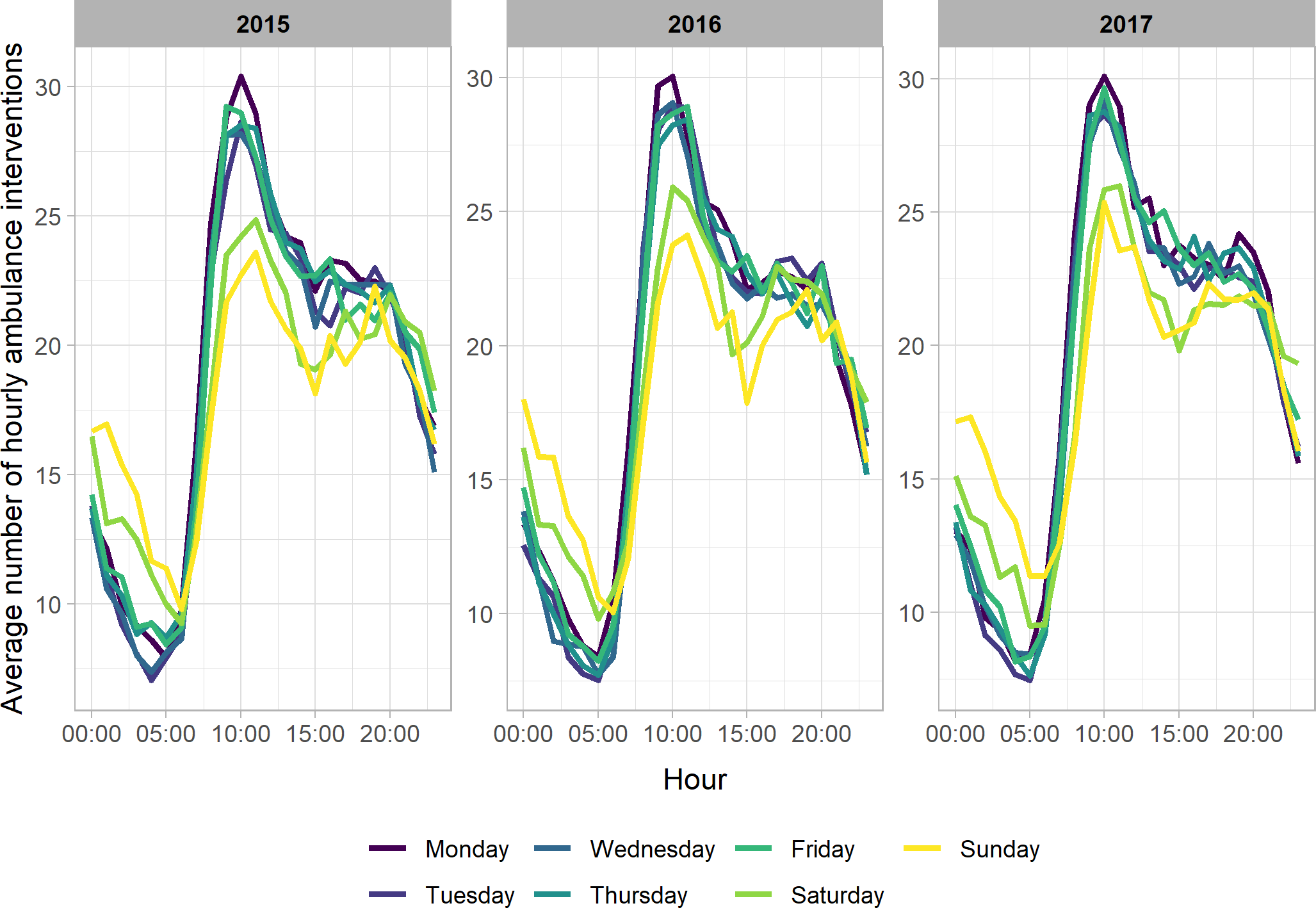}
  \caption{Average number of hourly ambulance interventions divided by the day of the week. There are clear seasonal patters that characterise weekends and weekdays.}
  \label{fig:hourly-events}
\end{figure}

Figure \ref{fig:hourly-events} displays the temporal dynamics of the emergency interventions within a day. The panel summarises the average number of hourly ambulance interventions splitted by the hour of the day and the day of the week. A similar pattern is found in the three years: after rapidly increasing in the early morning, the time series peaks around 10:00, slowly falls until 15:00, and remains stable until 20:00 when it rapidly drops until the next day. The hourly seasonalities are different between weekends and weekdays. In fact, possibly due to the city’s nightlife, the regional EMS registers, on average, more interventions during the first hours of the day at the weekend with respect to the rest of the week. Instead, lower frequencies are detected during the rest of the day.

\begin{figure}[tb]
  \centering
  \includegraphics[width=0.85\linewidth]{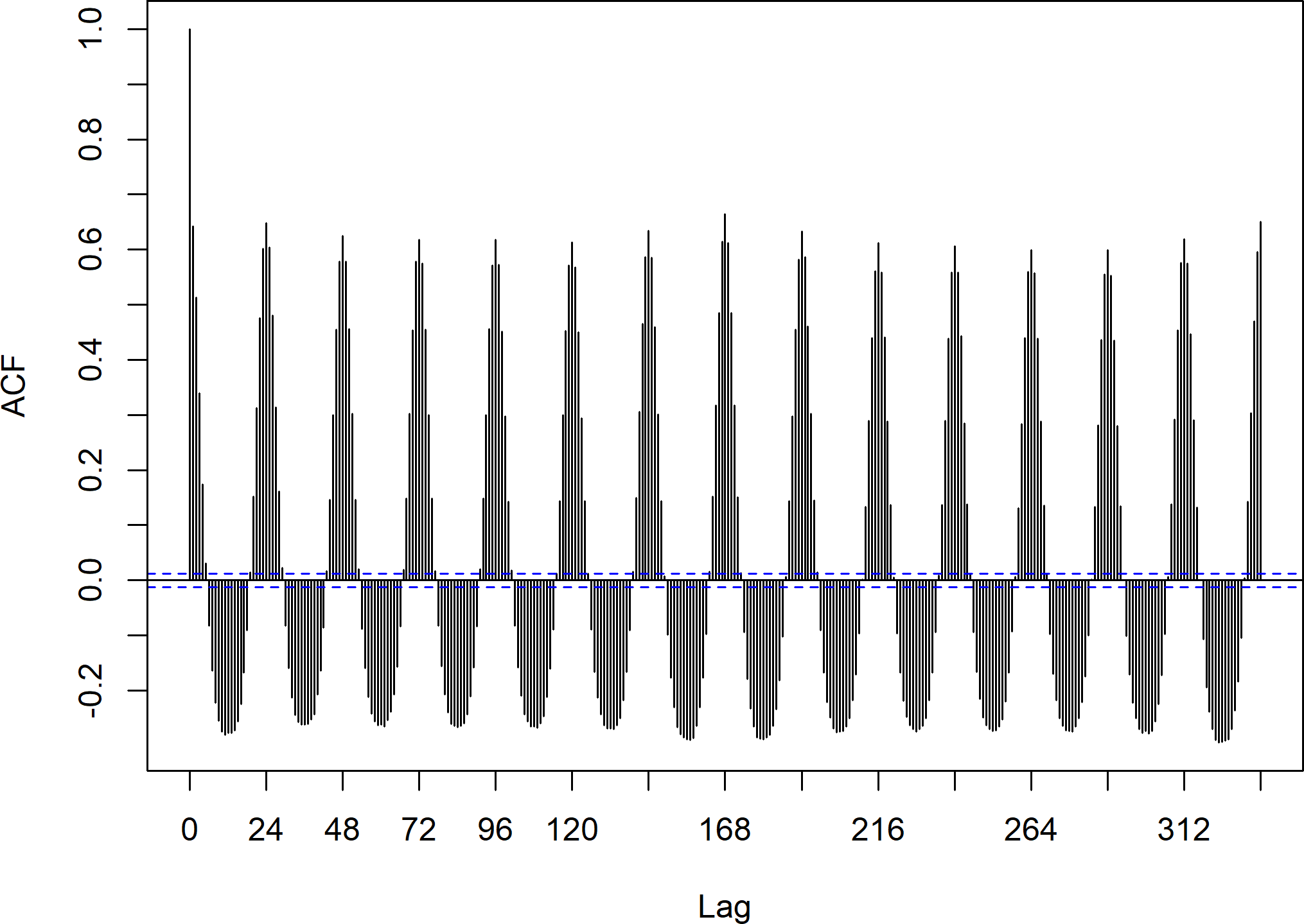}
  \caption{Auto-correlation function of the hourly number of EMS interventions occurred in the street network of Milan from 2015 to 2017. It clearly displays daily and weekly seasonalities.}
  \label{fig:ACF}
\end{figure}

We completed the temporal exploratory analysis examining the auto-correlation function (ACF) of the hourly number of EMS interventions. A graphical output considering two weeks of lagged counts is reported in Figure \ref{fig:ACF}. The plot clearly highlights hourly, daily, and weekly seasonalities. We can also notice that the ACF is negative when considering lags that are multiple of twelve, pointing out an opposite behaviour during mornings, afternoons, and nights. The statistical model proposed in Section \ref{sec:methods} takes into account the temporal patterns detected in the data. 

As already mentioned, we analysed the ambulance interventions as a spatial point process occurring on a restricted one-dimensional spatial domain which represents Milan's road network. More generally, a \emph{linear network}, denoted hereinafter by $L$, is defined as the union of a finite set of segments, say $l_i$, lying in a planar region $S$ \parencite{ang2012geometrically, baddeley2021analysing}
\[
l_i = [\mathbf{u}_i, \mathbf{v}_i] = \lbrace \mathbf{s} : \mathbf{s} = t\mathbf{u}_i + (1 - t) \mathbf{v}_i; \ 0 \le t \le 1 \rbrace,
\] 
where $\mathbf{u}_i$ and $\mathbf{v}_i$ denote the endpoints of $l_i$ stored using an appropriate Coordinate Reference System (CRS). In this paper, we adopt a projected CRS (EPSG code: 3003)\footnote{See \url{https://epsg.io/3003} for more details. Last access: 2023-03-06.} that expresses the units in metres.

The computational representation of Milan's road network adopted in this paper was created using data downloaded from Open Street Map (OSM) servers, and, in particular, we used the \emph{openstreetmap.fr}\footnote{See \url{http://download.openstreetmap.fr/}. Last access: 2021-12-09.} provider via the \texttt{R} package \texttt{osmextract} \parencite{R, osmextract}. OSM is a project that aims to build an open and editable map of the World \parencite{OpenStreetMap, barrington2017world}. The basic components of OSM data are called \emph{elements}, and they are divided into \emph{nodes},  which represent points on the earth's surface, \emph{ways}, which are ordered lists of nodes, and \emph{relations},  which are lists of nodes, ways and other relations where each member has additional information that describes its relationship with the other elements.

We downloaded OSM road data for Lombardia (the region of Northern Italy where Milan is located) and, using a spatial operation, we retained only the OSM elements that lay inside Milan's polygonal boundary. Then, we selected those segments that correspond to the most important streets of Milan, focusing on the following classes\footnote{We refer to \url{https://wiki.openstreetmap.org/wiki/Highways} for a comprehensive description of road network data in Open Street Map and guidelines for its classification system. We also refer to \url{https://wiki.openstreetmap.org/wiki/IT:Key:highway} for a comparison between the Italian classification system and the classes defined by OSM.} (listed in descending order of importance): \emph{motorways}, \emph{trunks}, \emph{primary}, \emph{secondary}, \emph{tertiary}, \emph{unclassified}, and \emph{residential}. We created a road network with $20064$ segments that longs approximately $1949$km, including the majority of the most important streets in Milan. 

The road network spreads all around the city and is depicted in Figure \ref{fig:spatial-distribution-4}. The white areas clearly identify suburb/non-urban places and some of the most iconic locations in Milan, like Parco Sempione, Giardini Indro Montanelli or City Life.

After creating the road network, we excluded all emergency calls whose GPS locations were found farther than 50 meters away from the closest segment of the network\footnote{The distance between a point and a segment was measured using the shortest euclidean perpendicular distance.}, since they probably occurred on other minor streets not included in the considered network. Approximately 14000 events were discarded, and the remaining interventions were projected to their nearest point of the network. The final sample included 480252 events. 

\begin{figure}[tb]
  \centering
  \includegraphics[width=0.9\linewidth]{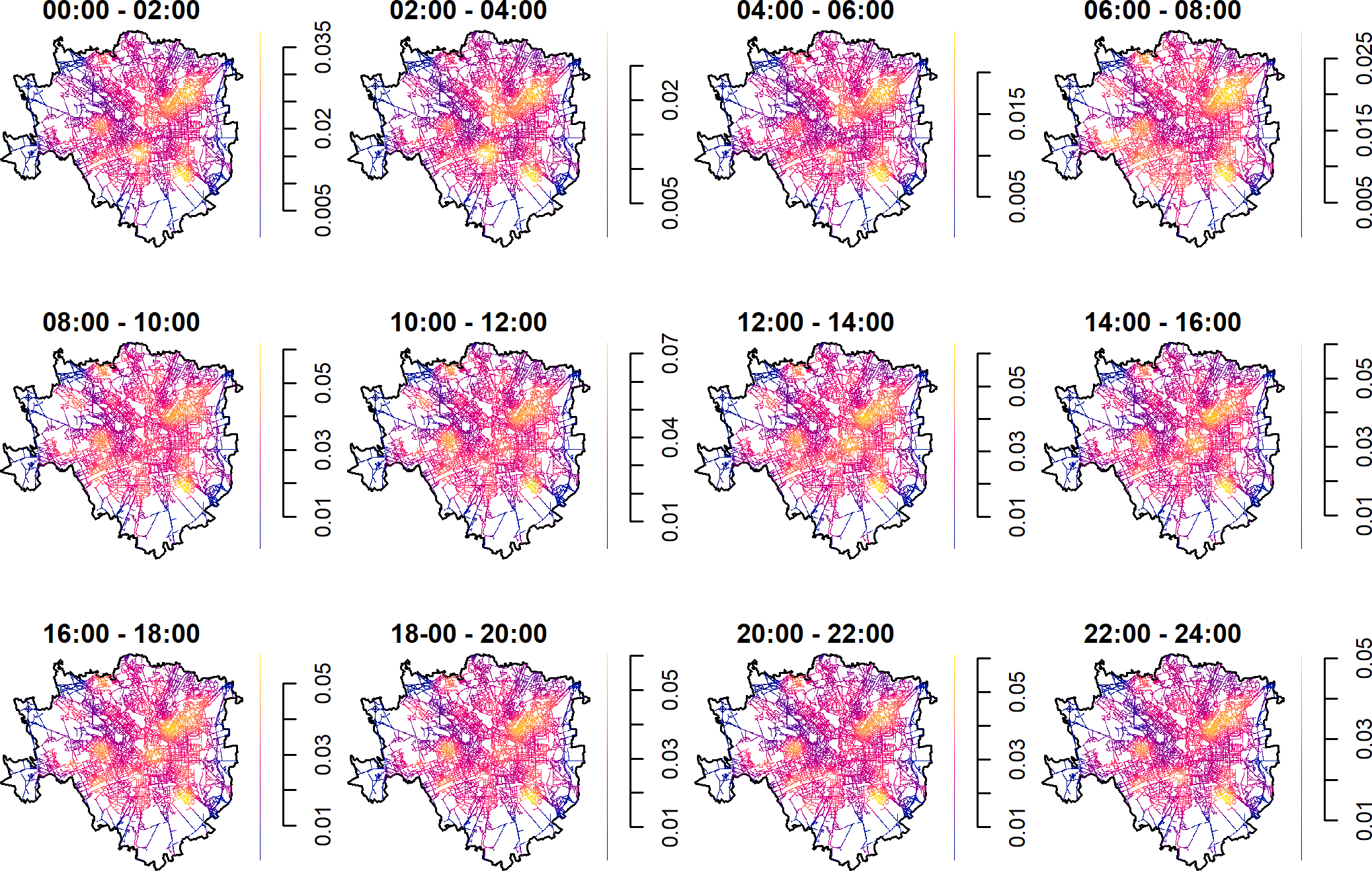}
  \caption{Smoothed intensity functions of EMS interventions in Milan estimated after classifying the events into two-hours classes.}
  \label{fig:hourly-spatial}
\end{figure}

Finally, we explored the spatio-temporal dynamics, testing the presence of space-time interactions. First, we split all EMS interventions into twelve two-hours classes according to their occurrence times. Then, we calculated (independently for each class) a smoothed intensity surface using the convolution kernel estimator detailed in \textcite{rakshit2019fast}. The result is reported in Figure \ref{fig:hourly-spatial}. We notice that from 10 AM to 06 PM the smoothed intensity peaks in the proximity of Duomo and the city centre, whereas, during the night hours, the interventions are concentrated in the proximity of night-life areas. In both scenarios, there are some clear hot spots near the main train station (i.e. Milano Centrale) and several nursing homes. The spatial model detailed in Section \ref{sec:methods} also includes these potential space-time interactions.

\section{Statistical model}
\label{sec:methods}

Let $L$ be a continuous one-dimensional spatial region and $\mathcal{T} = \lbrace 1, 2, \dots, T \rbrace$ a discrete temporal dimension divided into intervals of one hour. As mentioned before, in this paper $L$ represents the street network of Milan while $T$ is equal to 26304, i.e. the number of hours from 2015-01-01 00:00 to 2018-01-01 00:00. 

Let $y_t$ denote the number of ambulance interventions that occurred in the network $L$ at time $t \in \mathcal{T}$ and let $\bm{s}_{t, i}, \ i = 1, \dots, y_{t}$, denote the location of the $i$th event. We assume that, for each $t \in \mathcal{T}$, $\lbrace \bm{s}_{t, i} : i = 1, \dots, y_{t}\rbrace$ is a realisation of a \emph{non-homogeneous Poisson Process} (NHPP) on a linear network with \emph{intensity function} $\lambda_t(\bm{s})$ \parencite{okabe2012spatial, diggle2013statistical, baddeley2021analysing}. A NHPP on a linear network satisfies the following two properties:

\begin{itemize}[noitemsep]
	\item the number of events occurring in $L'\subseteq L$, which is denoted by $N(L')$, follows a Poisson distribution with parameter $\int_{L'} \lambda_t(\bm{s})\, \text{d}_1\bm{s}$, where $L'$ represents a finite portion of the network $L$ and $\text{d}_1\bm{s}$ denotes integration with respect to arc-length measure;
	\item let $N(L) = y_t$; then the $y_t$ events represent a random sample from a distribution whose probability density function is proportional to $\lambda_t(\bm{s})$.
\end{itemize}

We assume that the intensity function of the process can be decomposed as 
\begin{equation}
	\lambda_t(\bm{s}) = \mu_t g_t(\bm{s}) \quad \text{ for } \bm{s} \in L, 
	\label{eq:sep}
\end{equation}
where $\mu_t$ and $g_t(\bm{s})$ represent the temporal and spatial dimension at time $t$, respectively. We also assume that $g_t(\bm{s})$ satisfies the following two conditions
\begin{itemize}[noitemsep]
	\item $g_t(\bm{s}) > 0 \ \forall t \in \mathcal{T}$ and $\forall \bm{s} \in L$,
	\item $\int_{L} g_t(\bm{s}) \, \text{d}_1\bm{s} = 1 \ \forall t \in \mathcal{T}$,
\end{itemize}
which imply $\mu_t = \int_{L} \lambda_{t}(\bm{s}) \, \text{d}_1\bm{s}$. Therefore, considering the NHPP hypothesis, we notice that $y_{t} | \lambda_{t} \sim \text{Poisson}(\mu_t) $ $\forall t \in \mathcal{T}$, where $\mu_t$ represents the \emph{total volume} of ambulance dispatches at time $t$. Moreover, under the same assumptions, we have  $\bm{s}_{t,i} | \lambda_t, y_t \overset{\text{iid}}{\sim} g_t(\bm{s})$ for $i = 1, \dots, y_t$, highlighting that $g_t(\bm{s})$ denotes the spatial density of ambulance interventions at time $t$. Finally, we remark that our modelling scheme entails a temporal correlation among the interventions but it assumes that, for a fixed period $t$, the spatial locations are independent given $\lambda_t(\bm{s})$ and conform to an inhomogeneous Poisson process.

Equation \eqref{eq:sep}, despite being similar to the classical separability assumption for spatio-temporal point processes \parencite{moller2003statistical, diggle2013statistical}, suggests that the spatial component evolves over time. This particular functional form is motivated by the space-time interactions observed in the hourly evolution of ambulance interventions that were displayed in Figure \ref{fig:hourly-spatial}. In fact, as reported by several authors (see, e.g., \textcite{gonzalez2016spatio} and references therein), the separability of the first-order intensity function is usually taken as a working assumption in order to simplify the estimation process. However, given the exploratory analysis detailed before, we believe that this is not appropriate for our case. Therefore, following the ideas in \textcite{zhou2015predicting}, in this paper, we propose adopting a non-separable first-order intensity function readapted to analyse ambulance interventions as a point pattern on a linear network. As explained in Section \ref{sec:spat-mod}, the space-time interactions are modelled using an appropriate set of weights. 

Hereinafter, we introduce two statistical models for $\mu_t$ and $g_t(\bm{s})$, respectively. The temporal component is modelled using a semi-parametric Poisson regression with smoothed deterministic calendar covariates, namely the hour of the day, the day of the week, the week of the year, and an additional term allowing yearly fluctuations in the expected EMS counts. The spatial dimension is modelled non-parametrically using a network re-adaptation of a weighted kernel density estimator (KDE). 

\subsection{The temporal model}
\label{sec:temp-mod}

As mentioned above, the term $\mu_t$ represents the expected number of interventions that occurred over the network $L$ at time $t \in \mathcal{T}$. Following the suggestions in \textcite{diggle2005point, bayisa2020large}, we model $y_t$ using a Poisson regression. To incorporate smoothness into the model, Generalized Additive Models (GAMs) are used in the estimation of $\mu_t$ \parencite{wood2011fast}. GAMs extend Generalized Linear Models allowing for non-linear relationships between the response variable and the covariates. 

Being $y_t$ the observed number of emergency interventions in the linear network $L$ at time $t$, under a Poisson distribution assumption one has $\mu_t=E(y_t)$, and the log-linear Poisson additive regression model is given by
\[
\log(\mu_t) = \beta_0 +
\beta_1 \cdot \text{year}_t + \beta_2(\text{hour}_t) + \beta_3(\text{week}_t),
\]
where $\beta_0$ denotes the intercept and $\text{year}_t=0,1,2$  represents the year of the event occurred at time $t$ with respect to 2015. In addition, $\text{hour}_t$ represents the hour of the day (taking values from 0 to 23) while $\text{week}_t$ represents the week of the year (taking values from 1 to 53). The notation $\beta_j(x), \ j=2,3$ represent a spline transformation, i.e. $\beta_j(x)=\sum_{r=1}^{k_j}b_{jr}\gamma_{jr}(x)$, where $\gamma_{jr}(x), r=1,\ldots,k_j$ are the basis functions and $b_{jr}$ the unknown coefficients. In particular, a cyclic cubic regression spline is adopted since in our context it is appropriate to assume a smooth transition between the last hour of one day and the first hour of the next day as well as between the last week of one year and the first week of the next year (see Figures \ref{fig:daily-events} and  \ref{fig:hourly-events}).

To account for potential different impacts among the days of the week (see Figure \ref{fig:hourly-events}), an interaction term was also included in the linear predictor. Hence, the final model fitted to the data writes as follows
\begin{equation}
	\log \mu_t= \beta_0 + 
	\beta_1 \cdot \text{year}_t + \text{dow}_t + 
	\text{dow}_t \times \beta_2(\text{hour}_t) + 
	\beta_3(\text{week}_t).
	\label{eq:temp-mod}
\end{equation}
The term $\text{dow}_t$ is a factor variable that represents the day of the week.

\subsection{The spatial model}
\label{sec:spat-mod}

The spatial component of the intensity function, previously denoted by $g_{t}(\bm{s})$, is modelled non-parametrically using a network re-adaptation of a Jones-Diggle corrected weighted kernel density estimator \parencite{diggle1985kernel, jones1993simple, rakshit2019fast}. The weights are computed using a weight function that takes into account the space-time interactions described in Section \ref{sec:data}.

More precisely, given a set of observed time periods $\mathcal{T}$, an hour $u$, and a location $\bm{s}$ on the network, the weighted kernel estimator can be written as 
\begin{equation}
\hat{g}_{u}(\bm{s}) = \frac{\sum_{t \in \mathcal{T}} \sum_{i = 1}^{y_t} w(t, u) K_N(\bm{s}, \bm{s}_{t, i}; h)}{\sum_{t \in \mathcal{T}} \sum_{i = 1}^{y_t} w(t, u)},
\label{eq:spat1b}
\end{equation}
where $w(t, u)$ represents the weight associated to the $\bm{s}_{t, i}$ ambulance intervention and $K_N(\bm{s}, \bm{s}_{t, i}; h)$ denotes the Jones-Diggle corrected network kernel function. Following the proposal in \textcite{rakshit2019fast}, the kernel function is defined as
\begin{equation}
K_N(\bm{s}, \bm{s}_{t, i}; h) = \frac{K(\bm{s} - \bm{s}_{t, i}; h)}{c_L(\bm{s}_{i, t})},
\label{eq:K-N}
\end{equation}
where $K(\bm{s} - \bm{s}_{t, i}; h)$ denotes a planar Gaussian kernel with bandwidth $h$ and $c_L(\bm{s}_{i, t}) = \int_L K(\bm{s} - \bm{s}_{i, t}, h) \ \text{d}_1 \bm{s}$ represents the convolution of kernel $K$ with arc-length measure on the network. 

The KDE in Equation \eqref{eq:K-N} is one of the most relevant examples of statistical estimators for spatial network data that is based on euclidean distances instead of shortest path distances. For this reason, it can be expressed in terms of convolutions of kernels and can be computed extremely efficiently using the Fast Fourier Transformation (FFT) \parencite{silverman1982algorithm}. More precisely, the numerator in~\eqref{eq:K-N} can be expressed as the convolution of a kernel $K$ with respect to the counting measure on the data points, whereas the denominator can be expressed as a convolution with respect to arc-length measure on the network. In both cases, the estimators can be computed rapidly using the FFT after discretising the point pattern and the linear network via a fine pixel grid. 

We end this Section observing that although the suggested kernel approach does not consider shortest-path distances computed on the network, the structure of the spatial domain is still taken into account by the denominator in Equation~\eqref{eq:K-N}. Moreover, as discussed in \textcite{rakshit2019fast}, the proposed technique consistently estimates the intensity function of a point process on a linear network and its statistical efficiency is only slightly sub-optimal with respect to other approaches (see, e.g., \textcite{mcswiggan2017kernel}), whereas the computational advantages are enormous for large networks as the one considered in this paper.

\subsubsection{Defining the weight function}
\label{sec:weight-function}

The weight function $w(t, u)$ is used to capture the contribution of each past observation to predict future ambulance demand by taking advantage of EMS data temporal patterns to improve the forecasting of future interventions. It incorporates space-time interactions into the weighted kernel estimator, creating a non-separable structure in the spatio-temporal intensity $\lambda_t(\bm{s})$. In particular, we assumed that $w(t, u)$ can be modelled as a function of the time lag between $u$ and $t$, say $m = u - t$. The following functional form, firstly proposed by \textcite{zhou2015predicting}, was adopted  
\begin{equation}
	w(t, u) = w(u - t) = w(m) = \rho_1^{(m)} + \rho_2^{(m)} \rho_3^{\sin^2\left(\frac{\pi m}{24}\right)}\rho_4^{\sin^2\left(\frac{\pi m}{168}\right)}.
	\label{eq:weight}
\end{equation}
Equation~\eqref{eq:weight} includes a separate coefficient for each seasonal pattern displayed in Figures~\ref{fig:ACF}: $\rho_1$ captures the short-term dependence while $\rho_3$ and $\rho_4$ measure the daily and weekly seasonalities with a periodicity equal to $24$ and $24 \times 7 = 168$ hours, respectively. The coefficient $\rho_2$ represents a discount factor added to fade out the product of daily and seasonal terms, whereas the term $\pi$ denotes the constant value 3.1415\dots The four parameters are bounded between $0$ and $1$ to avoid (unrealistic) exponential growths.  Consequently, $w(m)$ takes values in $(0, 2)$, which also prevents negative weights that would potentially result in a negative kernel estimate of the density function.

Unfortunately, estimating $\rho_1, \dots, \rho_4$ in a full likelihood-based approach entails a non-trivial computational burden. Consequently, we implemented an algorithm firstly suggested in \textcite{zhou2015predicting}. As mentioned above, the weight function aims to grasp the time regularities of EMS interventions (also displayed in Figure~\ref{fig:hourly-spatial}), giving more importance to those events that occurred in the proximity of the seasonality peaks. Therefore, $w(m)$ should reflect the temporal dependency depicted by the ACF of the hourly number of EMS interventions, mirroring the behaviour displayed in Figure~\ref{fig:ACF} and assigning a negligible weight to those observations that, from a temporal perspective, are unlikely to be important for future predictions.

For this reason, after calculating the empirical hourly ACF up to lag $M$ and taking its positive part, denoted by $\text{ACF}^{+} = \max(0, \text{ACF})$, the parameters $\rho_1, \dots, \rho_4$ were estimated by minimising the following loss function 
\begin{equation}
\frac{1}{M}\sum_{m = 1}^{M}\left(\text{ACF}^{+}(m) - \rho_0w(m)\right)^{2} \quad \text{s.t.} \ 0 < \rho_j < 1 \ \forall j = 0, \dots, 4.
\label{eq:min}
\end{equation}
The coefficient $\rho_0$ represents a further discount factor without any practical interpretation. 
It is used to scale $w(\cdot)$ between $0$ and $1$, in order to make it consistent with the ordinate range of $\text{ACF}^{+}$.
In this paper, we choose $M = 672$, which represents four weeks of historical temporal data, whereas the minimisation problem was solved using the \textit{box-constrained} method implemented in the \texttt{R} function \texttt{optim()}, initialising all parameters at a random value between 0 and 1 \parencite{byrd1995limited, R}.

\section{Results}
\label{sec:results}

We now present the results obtained when estimating the spatial and temporal models described in Sections \ref{sec:temp-mod} and \ref{sec:spat-mod}. All procedures were implemented using the software \texttt{R} \parencite{R} and several contributed packages. More precisely, the smooth temporal components were estimated using the package \texttt{mgcv} \parencite{wood2017generalized}, while the network-version of the Gaussian weighted kernel is implemented in the package \texttt{spatstat} \parencite{baddeley2015spatial}. We fitted the temporal model, the weight function, and the spatial kernel using a laptop with an AMD Ryzen 5 3500U with Radeon Vega Mobile Gfx 2.10 GHz processor, four cores and 8GB of RAM. After downloading the OSM data, it took approximately 5 minutes to build the computational representation of the point pattern on the street network, 2 minutes to estimate the temporal model and the parameters in the weight function, and 3 minutes to compute the spatial kernel estimator considering two different future time periods. 

\subsection{The temporal component}
\begin{table}

\caption{Estimates of the effects obtained after fitting the model described in Equation \eqref{eq:temp-mod}. The reference category for the daily seasonal term is Sunday.}
\label{tab:summary-temp-mod}
\centering
\begin{tabular}[t]{lrrrr}
\toprule
& Estimate & Standard Error & z-value & pvalue\\
\midrule
(Intercept) & 2.8333 & 0.0044 & 648.9070 & <0.001\\
Year & 0.0116 & 0.0018 & 6.5792 & <0.001 \\ 
Monday & 0.0193 & 0.0057 & 3.3835 & <0.001\\
Tuesday & -0.0244 & 0.0058 & -4.2161 & <0.001\\
Wednesday & -0.0177 & 0.0058 & -3.0807 & 0.002\\
Thursday & -0.0137 & 0.0057 & -2.3937 & 0.017 \\
Friday & 0.0024 & 0.0057 & 0.4292 & 0.668 \\
Saturday & -0.0052 & 0.0057 & -0.9218 & 0.357 \\
\bottomrule
\end{tabular}
\end{table}

As detailed in Section \ref{sec:temp-mod}, the temporal component was estimated using a GAM with deterministic predictors representing yearly fluctuations and hourly, daily, and weekly seasonal components. The cyclic cubic spline terms are included to capture the smooth intra-day dynamics and the weekly temporal trends. The daily effects are taken into account by considering an interaction term and a set of dummy variables. The observed counts originally presented five missing values from 2015-04-01 at 00:00 to 2015-04-01 at 04:00. These values were imputed using a GAM as in Equation \eqref{eq:temp-mod} that was trained using the interventions until 2015-03-31 at 23:00.

The estimates of $\beta_0$, $\beta_1$, and $\text{dow}_t$ coefficients for the complete model are summarised in Table \ref{tab:summary-temp-mod}. The intercept represents the (logarithm of) the mean number of interventions per hour, $\beta_1$ is the annual variation in the EMS counts with respect to 2015, and the remaining set of coefficients represents the deviation from the reference level, i.e. Sunday. The estimates highlight the presence of a tiny but significant and  positively increasing trend in the hourly number of interventions per year. Moreover, we can notice that the behaviour of the ambulance interventions during the weekend looks quite different from the first days of the week (e.g. Monday to Thursday), with Monday being the day when the majority of interventions take place, whereas Friday and Saturday are found to be not significantly different from Sunday at the usual significance levels.

\begin{figure}
\centering
\subfloat[Intra-day effects \label{fig:temp-smooth-daily}]{
\includegraphics[width=0.45\linewidth]{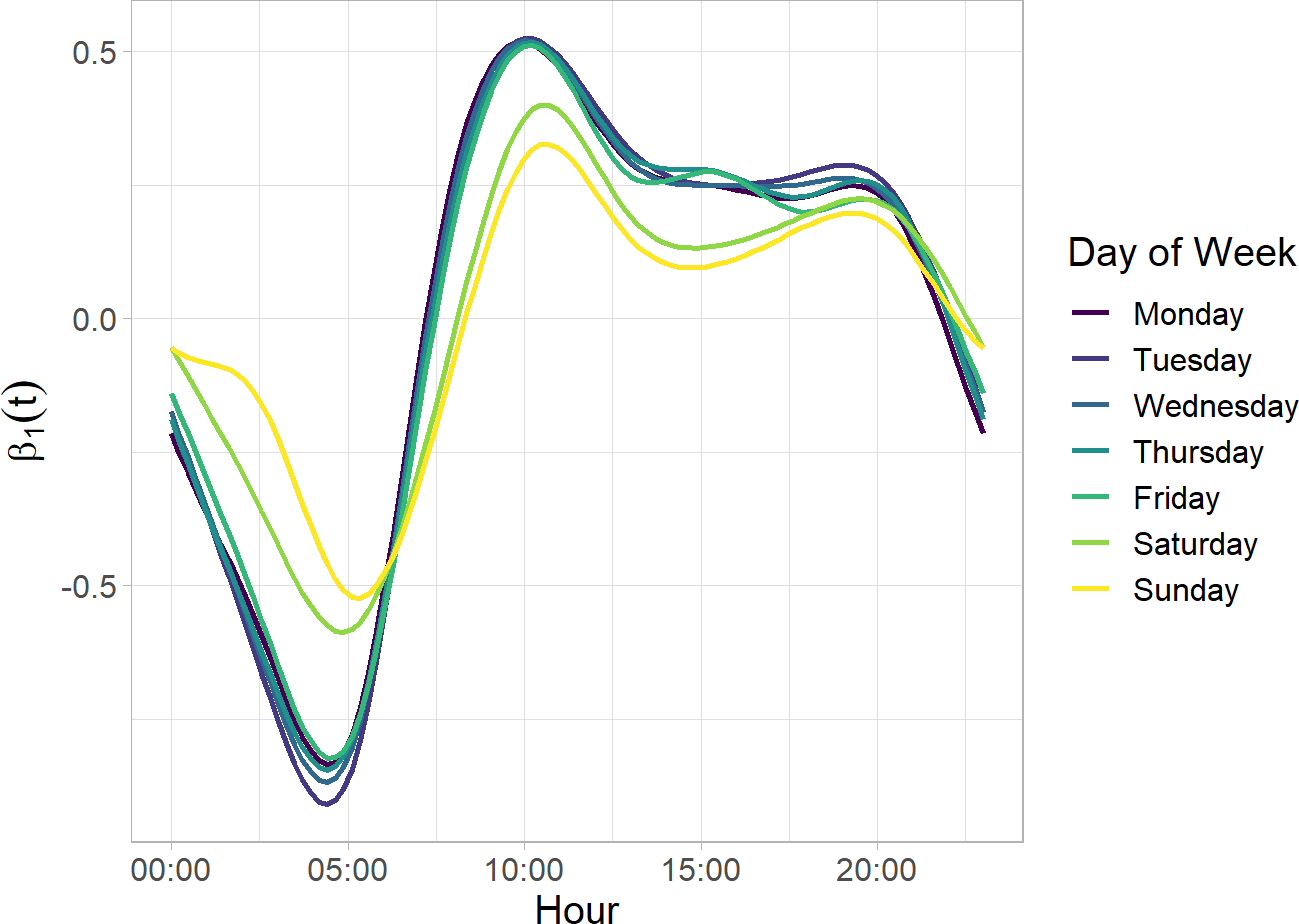}
}
\hspace*{0.25cm}
\subfloat[Weekly seasonality \label{fig:temp-smooth-weekly}]{
\includegraphics[width=0.45\linewidth]{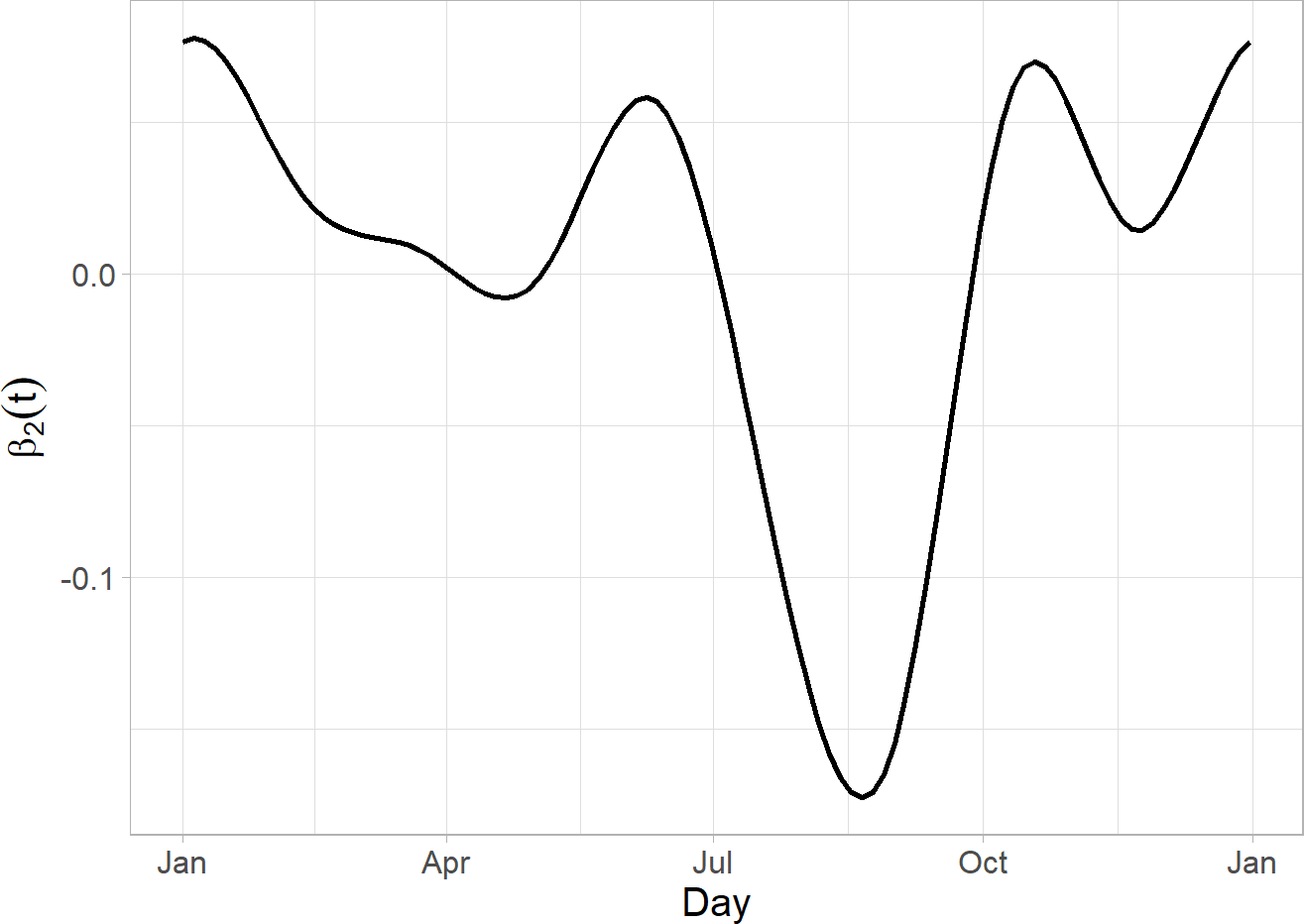}
}
\caption{Estimates of the smooth seasonal terms obtained after fitting the model described in Equation \eqref{eq:temp-mod}. Figure (a) represents the intra-day effects divided by the day of the week. Figure (b) displays the weekly temporal trends.}
\label{fig:temp-smooth}
\end{figure}

The smooth seasonal terms are depicted in Figure \ref{fig:temp-smooth}. In particular, Figure \ref{fig:temp-smooth-daily} reports the (smoothed) daily effects for each day of the week. The seven curves mirror the shapes displayed in Figure \ref{fig:hourly-events} and a clear distinction exists between weekends and weekdays. In all cases, we found a peak around 10 am. Figure \ref{fig:temp-smooth-weekly} shows the smoothed weekly effects, which look similar to the patterns displayed in Figure \ref{fig:daily-events}. We observe a drop in the expected number of ambulance dispatches around August. In both cases, the cyclic cubic regression splines were fitted using K = 10 knots evenly placed throughout the values of the covariates. 

\begin{figure}
\centering
\includegraphics[width = 0.85\linewidth]{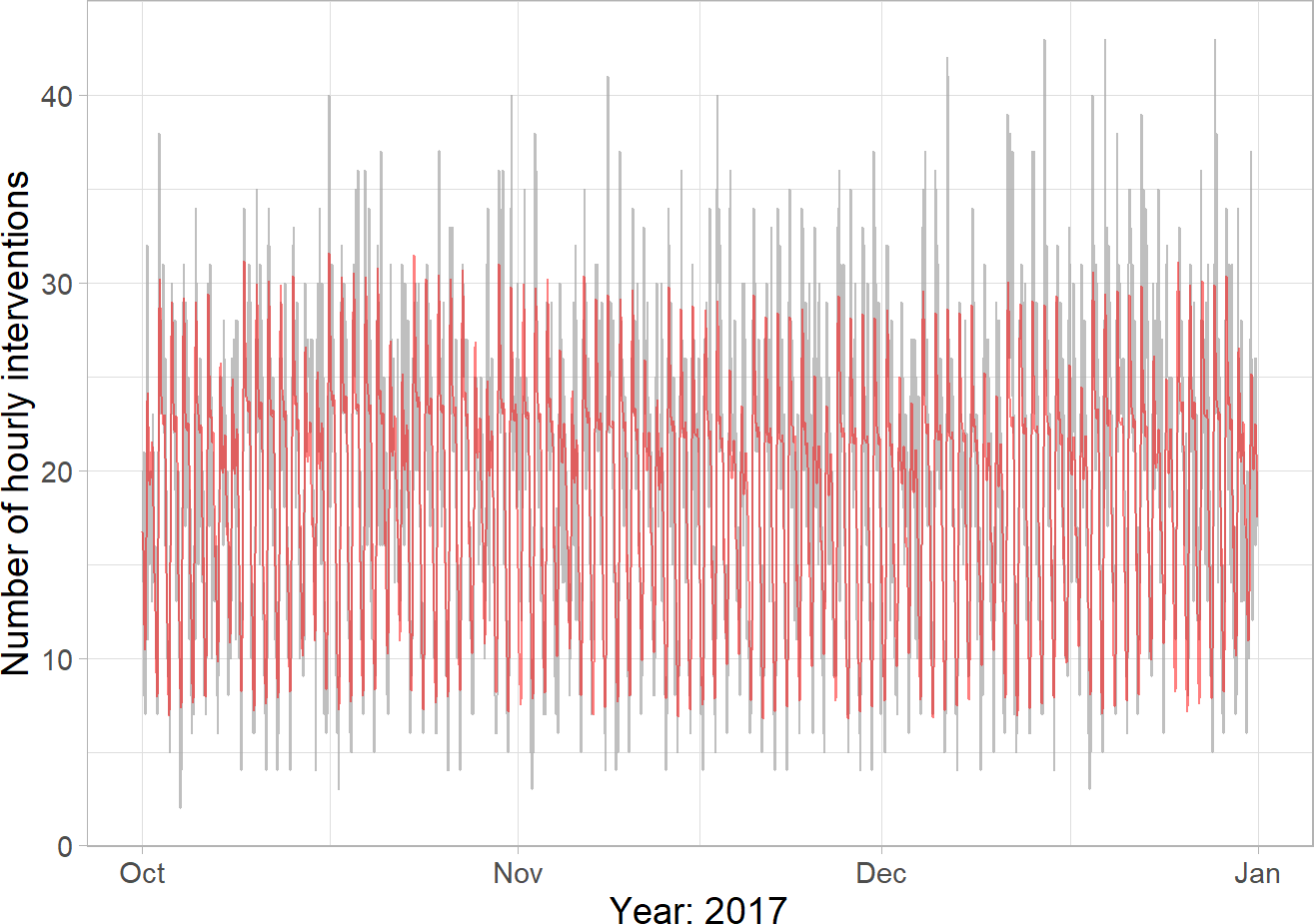}
\caption{Graphical comparison between observed counts (in grey) and out-of-sample fitted values (in red) considering EMS data from 2017-10-01 to 2017-12-31.}
\label{fig:temp-pred}
\end{figure}

Finally, we explored the predictive accuracy of the GAM using the following strategy. First, we trained the model considering all ambulance interventions up to 2017-10-01 at 00:00, and then we forecasted the EMS counts until the end of the year. We compared the observed counts with the out-of-sample fitted values and the result is displayed in Figure \ref{fig:temp-pred} that suggests a good agreement between the two time series.

\subsection{The spatial component}

As introduced in Section \ref{sec:spat-mod}, the spatial component was estimated combining a network re-adaptation of a Gaussian smoothing kernel with a weight function that measures the predictive importance of each past EMS intervention. The weights are used to mimic the interactions between two ambulance dispatches separated by $l$ temporal lags, replicating the hourly, daily, and weekly seasonalities in the ACF displayed in Figure \ref{fig:ACF}.

\subsubsection{Estimating the weight function}

As reported in Equation \ref{eq:weight}, the weight function depends on four parameters that represent the three seasonal components plus a discount factor. They were estimated solving the minimisation problem detailed in Equation \ref{eq:min}. We found $\rho_1$ as big as $0.213$, pointing out a mildly strong short-term correlation in the EMS counts. The second seasonal parameter, i.e. $\rho_3$, was found equal to $0.002$, which means that the effects related to the daily component range between $0.002$ and $1$. Given the periodic behaviour of the sinusoid function, the maximum value of $\hat{\rho}_3^{\ \sin^2\left(\frac{\pi l}{24}\right)}$ is obtained when the lag $l$ is approximately a multiple of 24, while the minimum is reached when the time difference is close to 12 or its odd multiples. The value of $\hat{\rho}_4$ was found equal to $0.927$, pointing out that the weekly effects are smoother and oscillate between $0.927$ and $1$. Finally, the fitted values of $\rho_0$ and $\rho_2$, i.e. the two discount factors, were found equal to $0.695$ and $0.999$, respectively, meaning that daily and weekly seasonality vanish slowly. 

\begin{figure}
\centering
\subfloat[One week of lagged counts \label{fig:ACFandweights1w}]{
\includegraphics[width=0.45\linewidth]{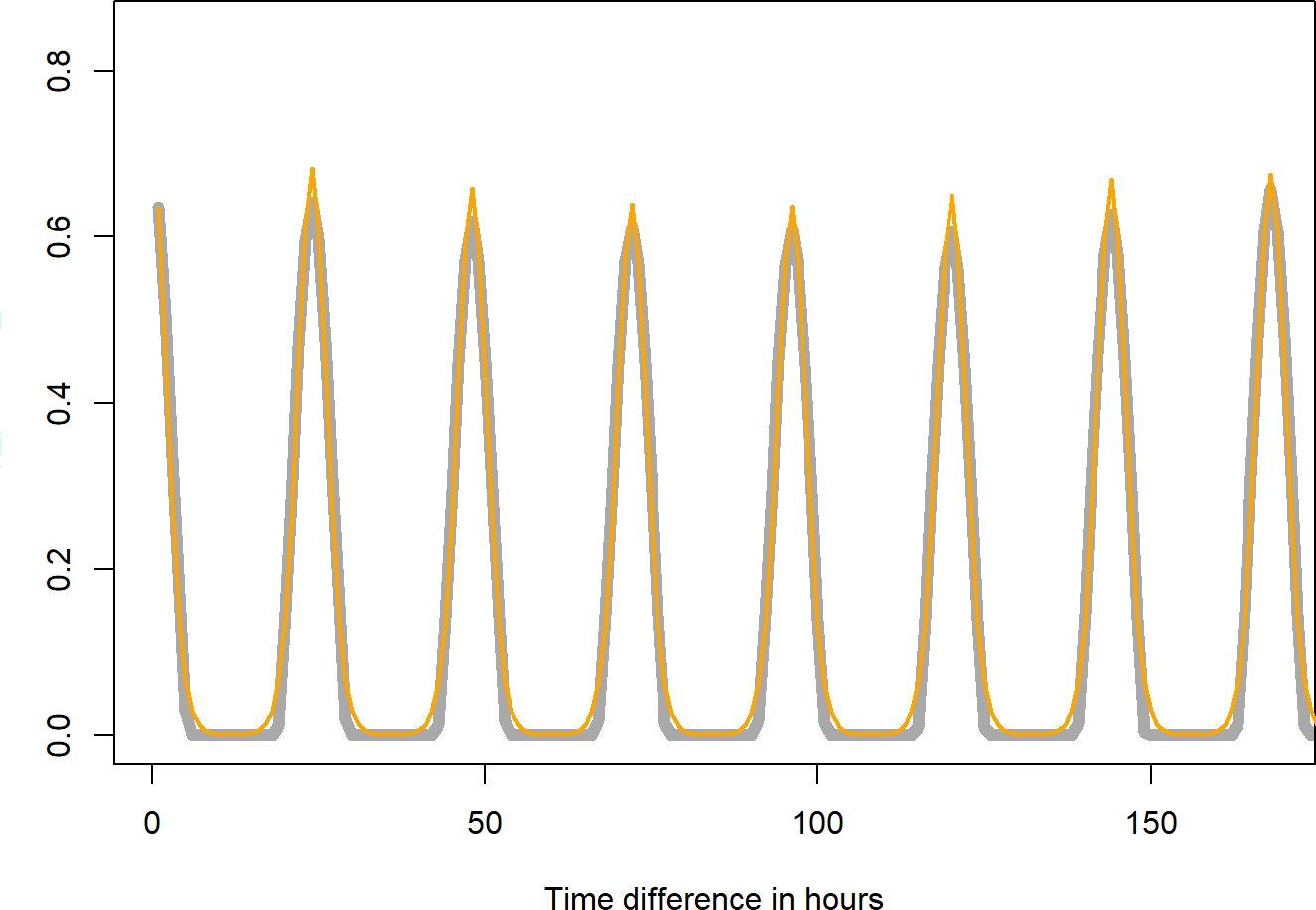}
}
\hspace*{0.25cm}
\subfloat[Four weeks of lagged counts \label{fig:ACFandweights4w}]{
\includegraphics[width=0.45\linewidth]{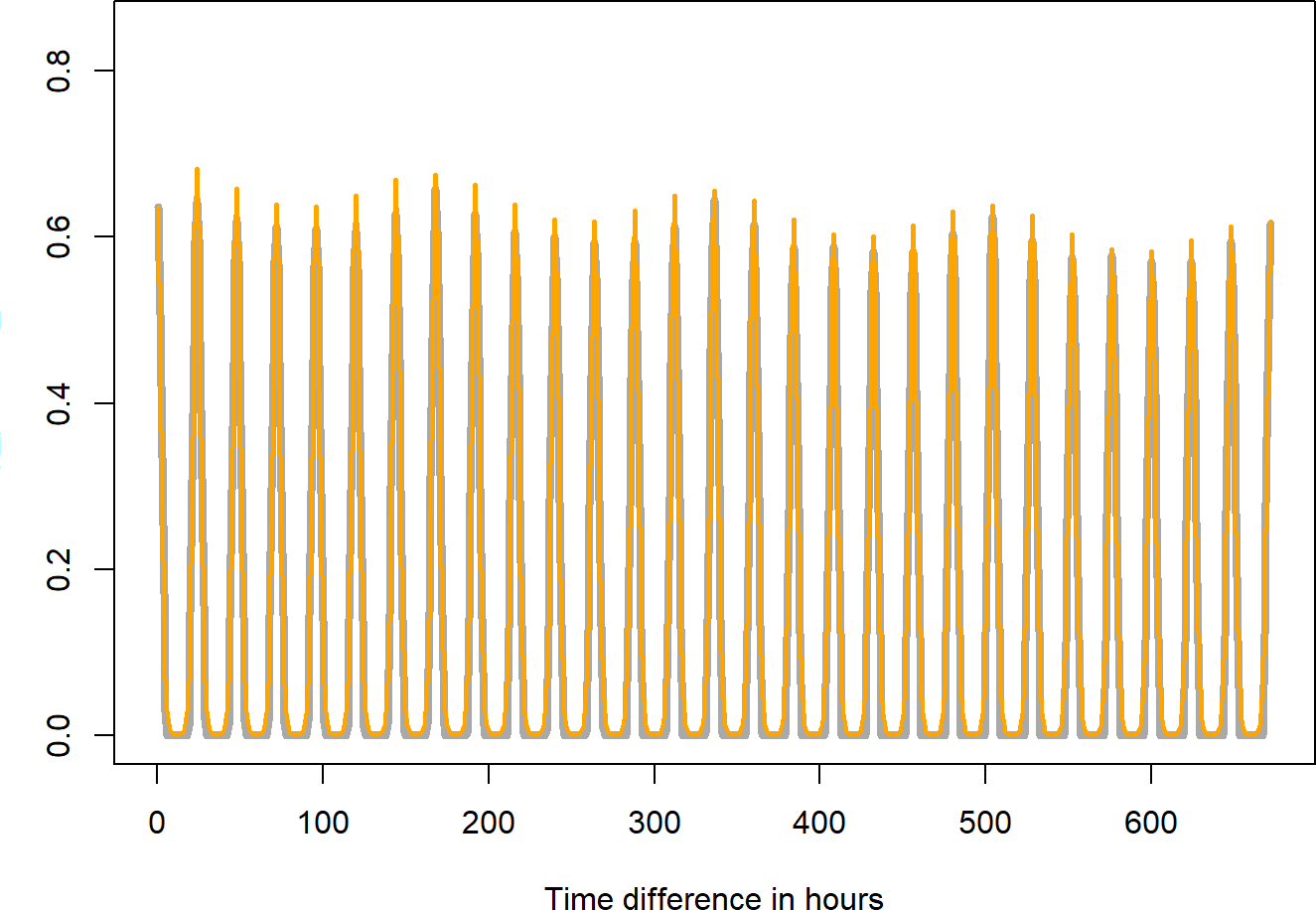}
}
\caption{The observed positive part of ACF (grey) and the estimated weight function (orange) considering lagged counts for (a) one week and (b) four weeks.}
\label{fig:ACFandweights}
\end{figure}

We display in Figure~\ref{fig:ACFandweights} a graphical comparison between the observed positive part of the hourly ACF and the estimates of the weight function. Figure~\ref{fig:ACFandweights1w} shows one week of lagged counts, while Figure~\ref{fig:ACFandweights4w} shows the complete set of lags up to four weeks. In both cases, the weight function successfully fits the ACF. 

\subsection{Spatial and spatio-temporal component}

After estimating the weight function, we applied Equation~\eqref{eq:spat1b} to obtain the predicted spatial density $\hat{g}_{u}(\bm{s})$ for a time period $u$. In particular, considering that the data at hand included the EMS interventions from 2015-01-01 at 00:00 to 2017-12-31 at 23:59, we decided to forecast $\hat{g}_{u}(\bm{s})$ considering two randomly selected future time periods: 2018-01-03 at 03:00 and 2018-01-03 at 15:00. The first one falls at night, while the other one falls in the early afternoon. In both cases, the value of the bandwidth $h$ was chosen using the rule of thumb suggested by \textcite[Section~8]{rakshit2019fast}. The bandwidth is given by 
\begin{equation}
h = (3n) ^ {-1/5}\bar{s},
\label{eq:scottrot}
\end{equation}
where $\bar{s} = \sqrt{s_1^2 + s_2^2}$ and $s_j, \, j = 1, 2, $ denotes the sample standard deviation of the $j$-th Cartesian coordinate values for the locations of the ambulance interventions. Equation~\eqref{eq:scottrot} adapts the rule of thumb proposed by \textcite[p. 152]{scott1992multivariate} in kernel density estimation to the analysis of spatial data lying on a 1-dimensional domain.

\begin{figure}
\centering
\subfloat[2018-01-03 03:00 \label{fig:pred-spat-dens-03}]{
\includegraphics[width=0.45\linewidth]{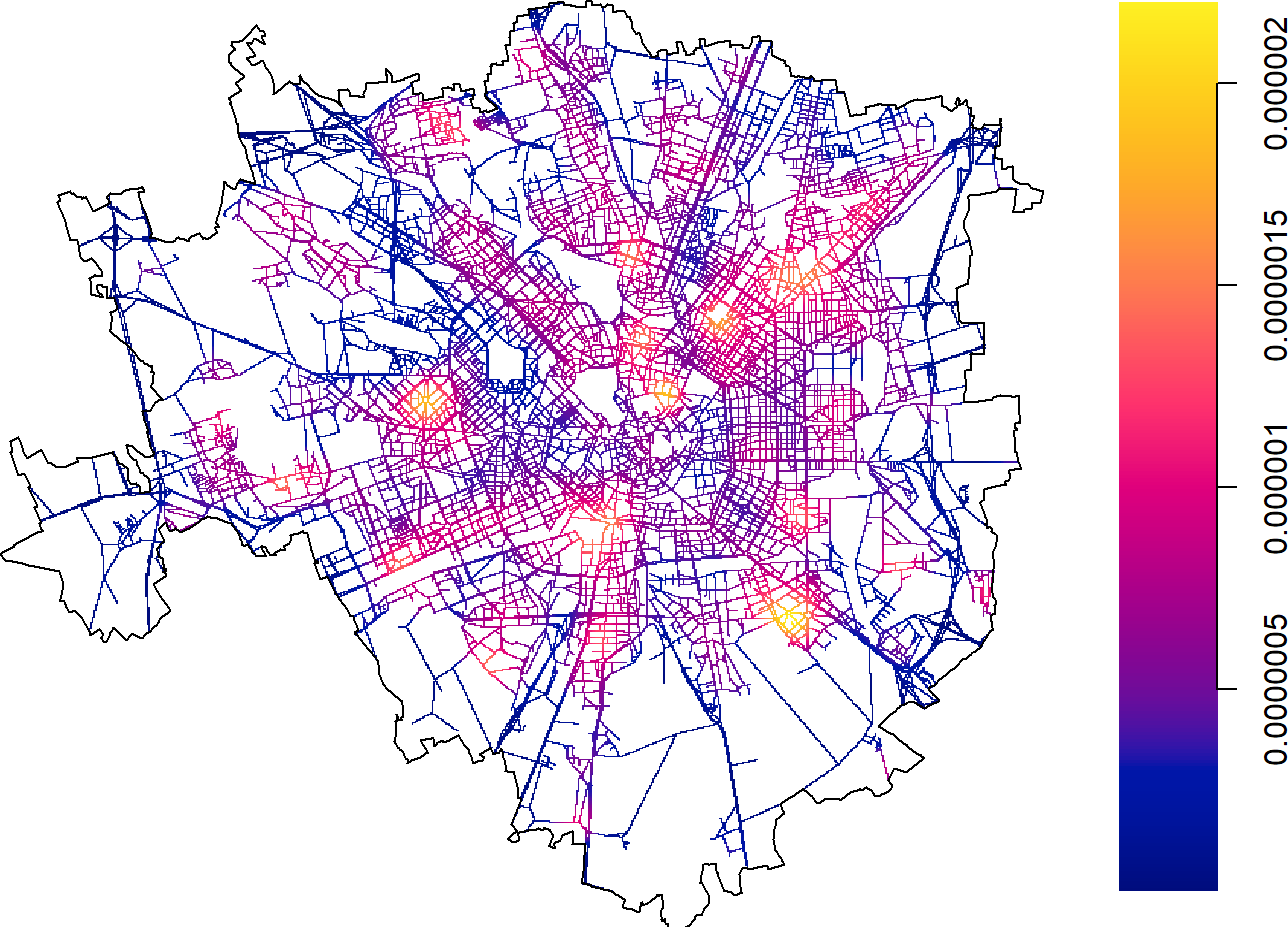}
}
\hspace*{0.25cm}
\subfloat[2018-01-03 15:00 \label{fig:pred-spat-dens-15}]{
\includegraphics[width=0.45\linewidth]{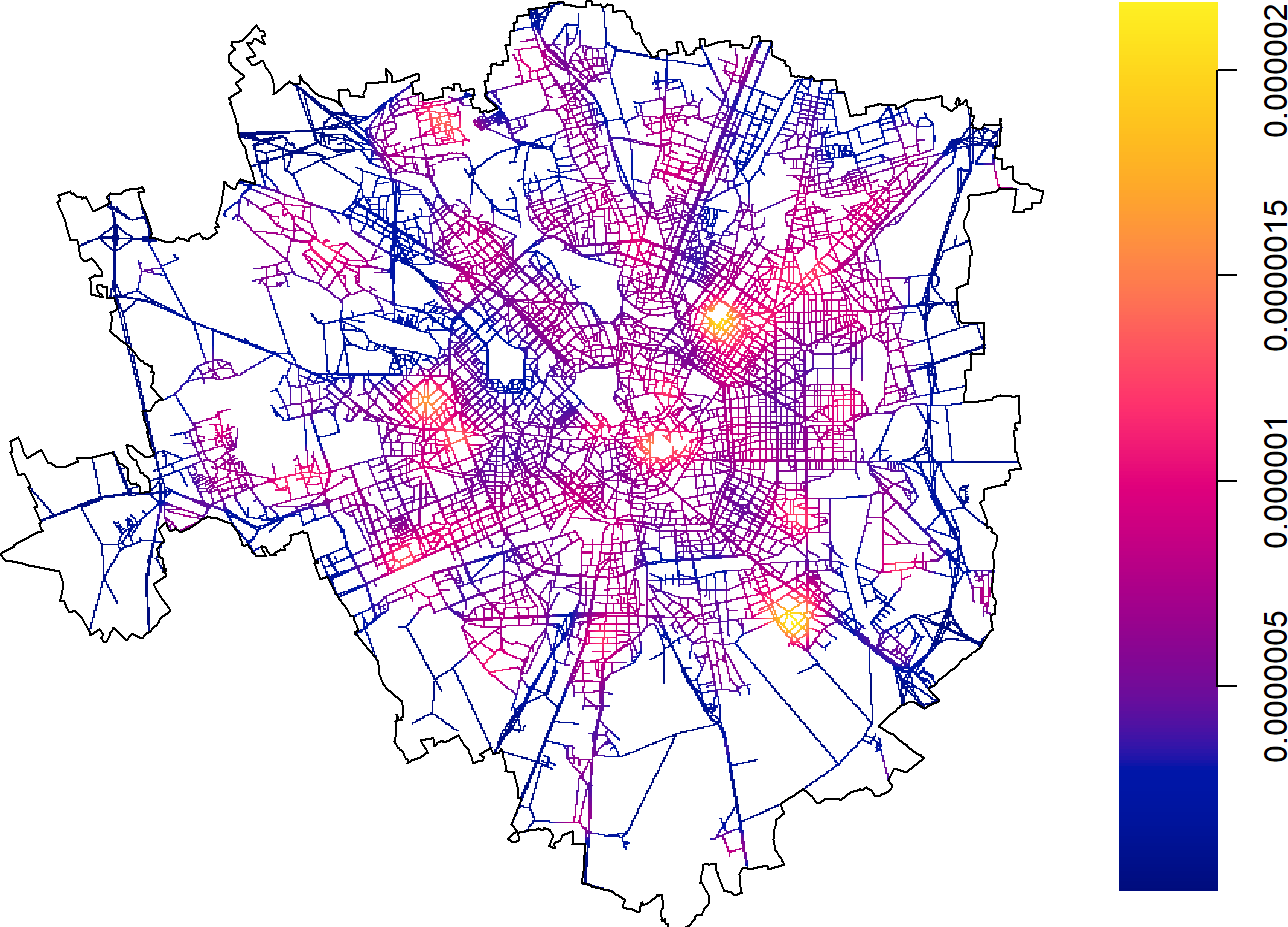}
}
\caption{Estimates of spatial density function $\hat{g}_{u}(\bm{s})$ considering two future time periods: 2018-01-03 at 03:00 (a) and 2018-01-03 at 15:00 (b). The unit for the colour scale is 1/m.}
\label{fig:pred-spat-dens}
\end{figure}

The results are reported in Figure~\ref{fig:pred-spat-dens}. Figure~\ref{fig:pred-spat-dens-03} shows that during the night the EMS interventions are spread in several parts of the municipality and highlights some roads of the network nearby night-life areas. Figure~\ref{fig:pred-spat-dens-15} underlines that ambulance dispatches are concentrated in the areas close to Duomo and other relevant working places during daytime, whereas night-life areas are no longer highlighted by the model. In both cases, the central station, a popular square (Piazzale Corvetto), and several retirement houses (such as Pio Albergo Trivulzio) are pointed out.

\begin{figure}
\centering
\subfloat[2018-01-03 03:00 \label{fig:pred-spat-int-03}]{
\includegraphics[width=0.45\linewidth]{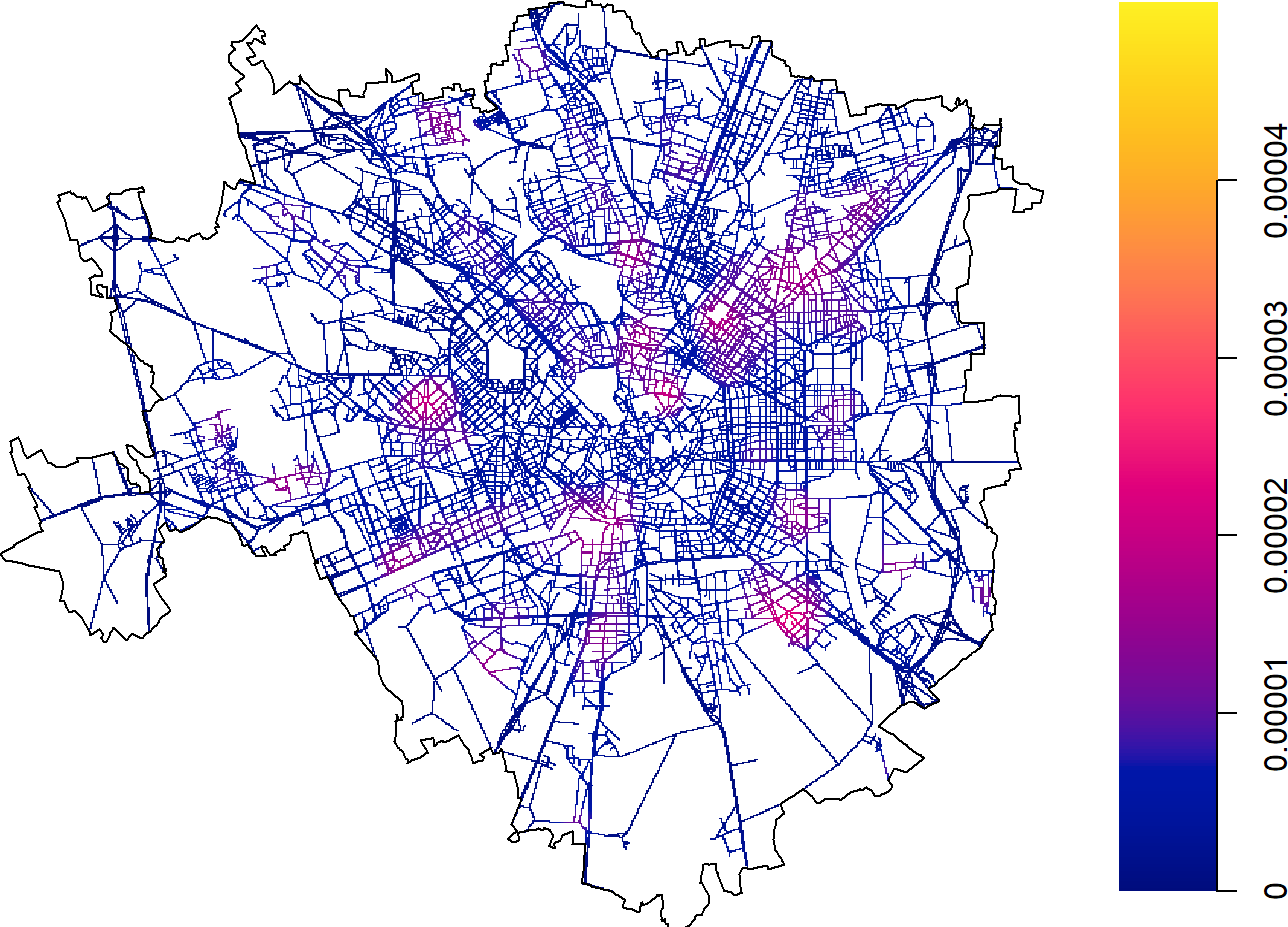}
}
\hspace*{0.25cm}
\subfloat[2018-01-03 15:00 \label{fig:pred-spat-int-15}]{
\includegraphics[width=0.45\linewidth]{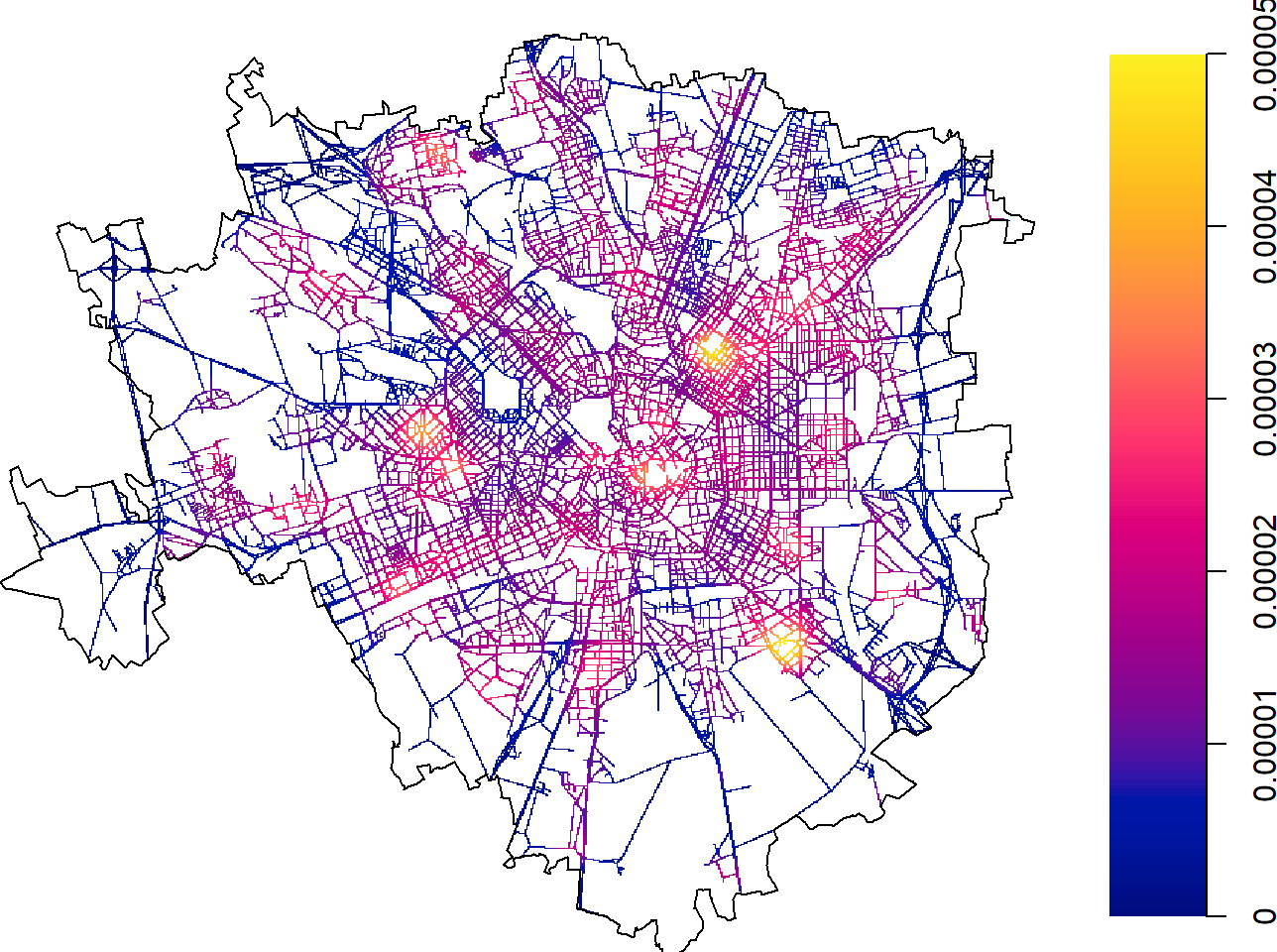}
}
\caption{Estimates of the intensity function $\hat{\lambda}_{u}(\bm{s})$ considering two future time periods: 2018-01-03 at 03:00 (a) and 2018-01-03 at 15:00 (b). The two maps highlight the temporal patterns in spatial locations of emergency interventions. The unit for the colour scale is 1/m. The values represent the expected number of ambulance dispatches occurring in a small linear neighbourhood around a point of the network.}
\label{fig:pred-spat-int}
\end{figure}

The values of $\hat{g}_{u}(\bm{s})$ displayed in Figure~\ref{fig:pred-spat-dens} represent only the spatial dimension of the data. Hence, to compare and visualise the spatio-temporal evolution of $\lambda_u(\bm{s})$, we estimated the expected number of interventions at time $u$, and calculated $\hat{\lambda}_u(\bm{s})$ by multiplying the spatial and the temporal components. The results are depicted in Figure~\ref{fig:pred-spat-int}. Although the two maps highlight areas similar to those displayed in Figure~\ref{fig:pred-spat-dens}, they now account for the temporal patterns of EMS interventions. In particular, considering that the majority of ambulance dispatches occur between 8 AM and 6 PM, the intensity function at 15:00 was found higher than in the other scenario.

\begin{figure}
	\centering
	\subfloat[2018-01-03 03:00 \label{fig:pred-spat-dens-se-01}]{
		\includegraphics[width=0.45\linewidth]{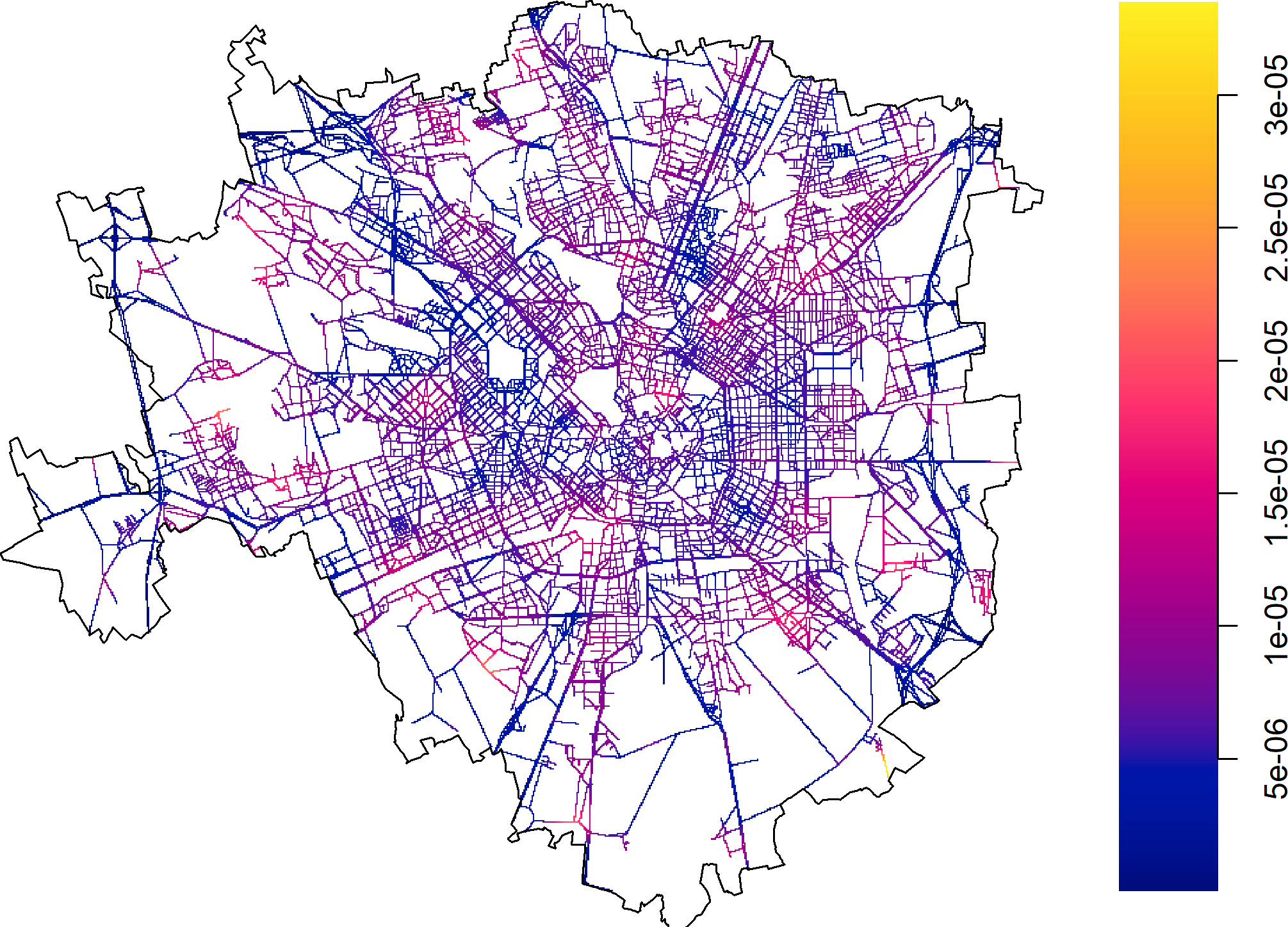}
	}
	\hspace*{0.25cm}
	\subfloat[2018-01-03 15:00 \label{fig:pred-spat-dens-se-02}]{
		\includegraphics[width=0.45\linewidth]{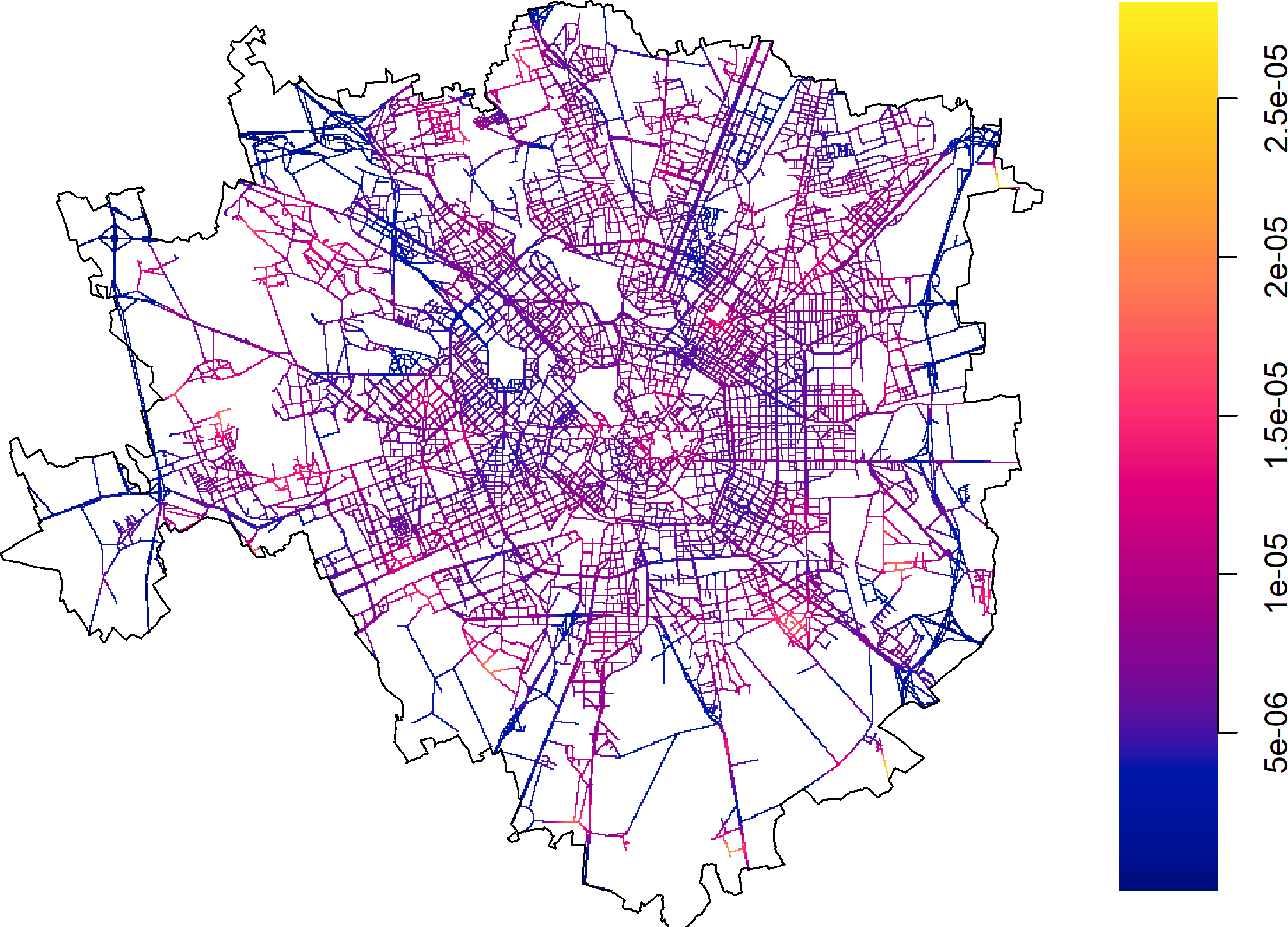}
	}
	\caption{Estimates of the standard errors of the spatial density function $\hat{g}_u(\bm{s})$ considering two future time periods: 2018-01-03 at 03:00 (a) and 2018-01-03 at 15:00 (b). The values were obtained using the formulas described in \textcite[Sec. 6.2]{rakshit2019fast}.}
	\label{fig:g-se}
\end{figure}

Finally, using the approach described in \textcite[Sec. 6.2]{rakshit2019fast}, we estimated the pointwise standard errors of the spatial density function $\hat{g}_u(\bm{s})$ considering the same two future time periods. The results are reported in Figure~\ref{fig:g-se}. In both cases, the two maps highlight certain areas of the municipality in the proximity of nightlife neighbourhoods or train and metro stations. We can also clearly recognise the shape of some arterial thoroughfares (e.g. the A51 motorway located on the rightmost part of the maps and far from any residential neighbourhood) passing through the city. Unsurprisingly, the standard error estimates generally increase in the proximity of the city boundary since they are based on fewer data points.

\section{Spatial validation}
\label{sec:validation}

Section \ref{sec:results} presented the results obtained when fitting the spatial and temporal components of $\lambda_t(\bm{s})$ and omitted any consideration on the spatial predictive accuracy. Nevertheless, as mentioned in the Introduction, the algorithms used to minimise the ambulance response times require a model that can produce reliable forecasts of the spatial distribution of the EMS events at the road network level. In this section, we discuss the procedure adopted to validate our proposal.

The predictive accuracy of $\hat{g}_t(\bm{s})$ was firstly inspected graphically by comparing observed and predicted EMS interventions at different levels of temporal aggregation via a network-readaptation of the \emph{relative-risk} function \parencite{mcswiggan2020estimation}. More precisely, given two point patterns $\bm{A}$ and $\bm{B}$ that occur on the same network $L$ and a point $\bm{s} \in L$, we define the \emph{(normalised) relative-risk function} (also named \emph{probability distribution of one type} in \textcite[Chapter~14]{baddeley2015spatial}) as
\[
\rho(\bm{s}) = \frac{g_{\bm{A}}(\bm{s})}{g_{\bm{A}}(\bm{s}) + g_{\bm{B}}(\bm{s})}, 
\]
where $g_{\bm{A}}(\bm{s})$ and $g_{\bm{B}}(\bm{s})$ denote the spatial densities of $\bm{A}$ and $\bm{B}$, respectively. The plug-in estimator of $\rho(\bm{s})$ is given by
\begin{equation}
\hat{\rho}(\bm{s}) = \frac{\hat{g}_{\bm{A}}(\bm{s})}{\hat{g}_{\bm{A}}(\bm{s}) + \hat{g}_{\bm{B}}(\bm{s})},
\label{eq:relriskplug}
\end{equation}
where $\hat{g}_{\bm{A}}(\bm{s})$ and $\hat{g}_{\bm{B}}(\bm{s})$ are kernel estimates of $g_{\bm{A}}(\bm{s})$ and $g_{\bm{B}}(\bm{s})$, respectively. The value of $\hat{\rho}(\bm{s})$ represents the probability that a point $\bm{s}\in L$ belongs to $\bm{A}$ instead of $\bm{B}$. Values of $\hat{\rho}(\bm{s})$ around 0.5 highlight that the relative risk function cannot discern the two processes. We refer to \textcite{mcswiggan2020estimation} for more details and extensive theoretical and computational considerations regarding the estimation of the relative risk for point patterns on linear networks.

As mentioned before, the forecasting accuracy of $\hat{g}_t(\bm{s})$ was tested by comparing observed points and (out of sample) predictions. More precisely, we first selected the EMS interventions that occurred before the end of September 2017 and trained the weight function to estimate $\rho_0, \dots, \rho_4$. Then, considering the temporal evolution of the weights, we derived the spatial KDE $\hat{g}_u(\bm{s})$ for each hour $u$ of a given (out-of-sample) time period $\mathcal{U}$, and, finally, we obtained an out-of-sample prediction of the emergency events by sampling $y_u$ points from each density $\hat{g}_u(\bm{s}), \ u \in \mathcal{U}$. Finally, we aggregated observed occurrences and predicted points over $\mathcal{U}$ and compared the two types of events by means of the relative risk function. The pseudo-code that summarises this procedure is reported in Algorithm \ref{alg:alg1}. 

In the analysis reported below, we tested the spatial accuracy considering four days placed farther and farther in time from the end of the training period, namely 2017-10-01, 2017-10-08, 2017-10-15, and 2017-10-22. We decided to focus on several days spread over a month close to the end of the training set since that represents a realistic scenario to organise the ambulance shifts. 

\begin{algorithm}[tb]
\SetAlgoLined
\SetKwFunction{subset}{subset}
\KwIn{\emph{data}: ambulance interventions data; $\mathcal{U}$: set of future time periods.}
\KwOut{An estimate of the normalised relative risk function.}
\tcc{1. subset EMS data that occurred before 2017-10-01 at 00:00.}
\emph{train} $\leftarrow$ subset(\emph{data}, \emph{date-occurrence} < 2017-10-01 00:00)\;
\tcc{2. estimate $\rho_0, \dots, \rho_4$ using the methods described in Section \ref{sec:weight-function}}
$\lbrace\hat{\rho}_0, \dots, \hat{\rho}_4 \rbrace\ \leftarrow$ estimate\_coefs(\emph{train})\;
\For{each $u \in \mathcal{U}$}{
\tcc{3. Estimate the weights $w(t, u)$ given $\hat{\rho}_0, \dots, \hat{\rho}_4$. See Equation \eqref{eq:weight}.}
$w$ $\leftarrow$ estimate\_weights($u$, $\hat{\rho}_0, \dots, \hat{\rho}_4$)\;
\tcc{4. Estimate $\hat{g}_u(\bm{s})$ using Equation \eqref{eq:spat1b}.}
$\hat{g}_u \leftarrow$ estimate\_KDE(\emph{ems\_train}, $w$)\;
\tcc{5. Simulate $y_u$ events sampling from a probability density function on a linear network equal to $\bm{g}_u$}
\emph{pred\_events(u)} $\leftarrow$ simulate\_points($\hat{g}_u$, $y_u$)
}
\caption{Pseudo code describing the procedure used to simulate future events for a non-separable model. The algorithm can be adapted to sample from a separable model (see Section~\ref{sec:compare}) skipping steps 2 and 3. In both cases, after the for loop, we aggregate all predicted events collapsing the temporal dimension.}
\label{alg:alg1}
\end{algorithm}

After simulating the ambulance interventions for each time period and extracting the corresponding observed EMS events, the relative risk function was computed as in \textcite{mcswiggan2020estimation}. The same bandwidth $h$ was used when applying Equation~\eqref{eq:spat1b} to the two types of points and its value was estimated using a re-adaptation of Scott's rule of thumb for one-dimensional coordinates data, as suggested in \textcite[pp.~5]{mcswiggan2020estimation}. We did not explore the other techniques for bandwidth selection due to the prohibitive computational costs of applying leave-one-out cross-validation to large road networks with hundreds of thousands of points and time-consuming out-of-sample simulations.

\begin{figure}
\centering
\includegraphics[width = 0.8\linewidth]{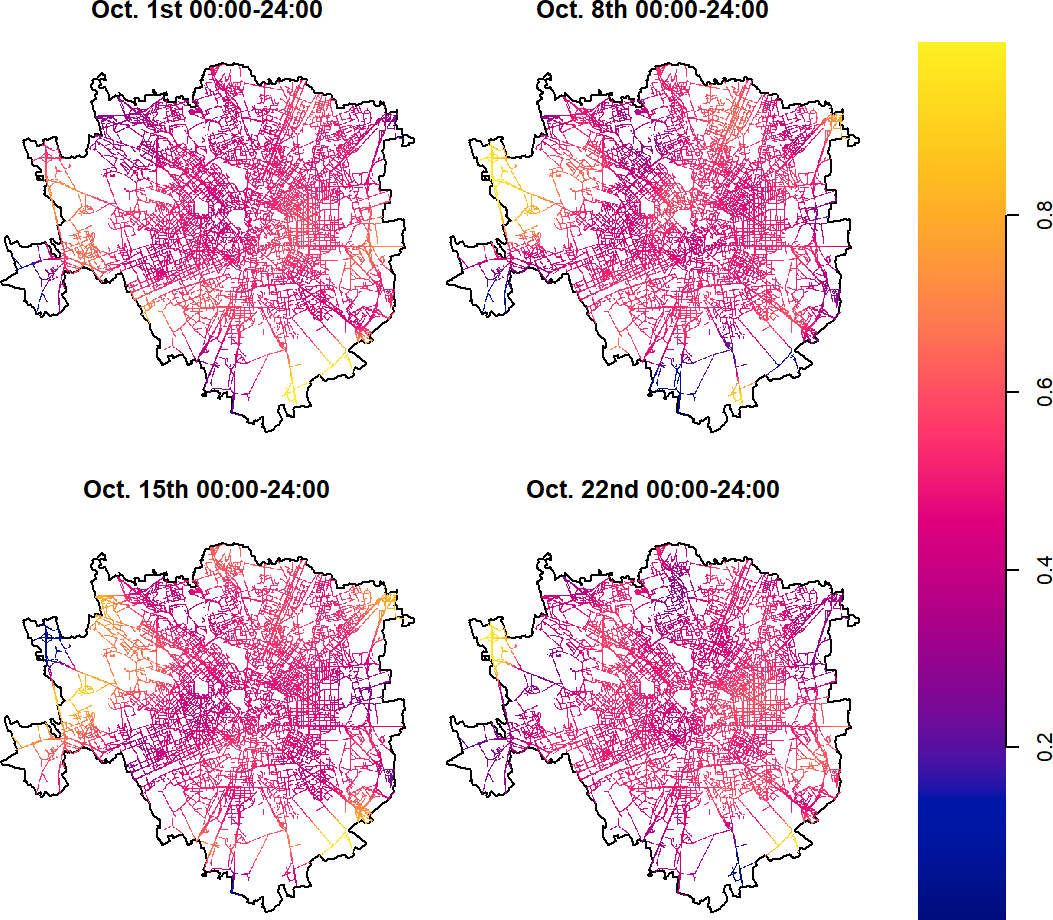}
\caption{Spatial representations of the (normalised) relative risk function considering four different days.}
\label{fig:relrisk-maps}
\end{figure}

The relative risk functions $\hat{\rho}(\bm{s})$ for the four days under analysis are depicted in Figure~\ref{fig:relrisk-maps}. Following the notation adopted in Equation~\eqref{eq:relriskplug}, the object $\bm{A}$ denotes the observed EMS events, while $\bm{B}$ represents the predicted points. A spatio-temporal EMS model successfully predicts future emergency events when $\hat{\rho}(\bm{s})$ is close to 0.5, since that implies the relative risk function cannot distinguish between observed and predicted cases. Figure~\ref{fig:relrisk-maps} shows that in the four cases the relative risk functions are always concentrated around 0.5 but for a few parts in the suburban areas, suggesting that our approach can be employed for EMS events forecasting. 

\begin{figure}
\centering
\includegraphics[width=0.85\linewidth]{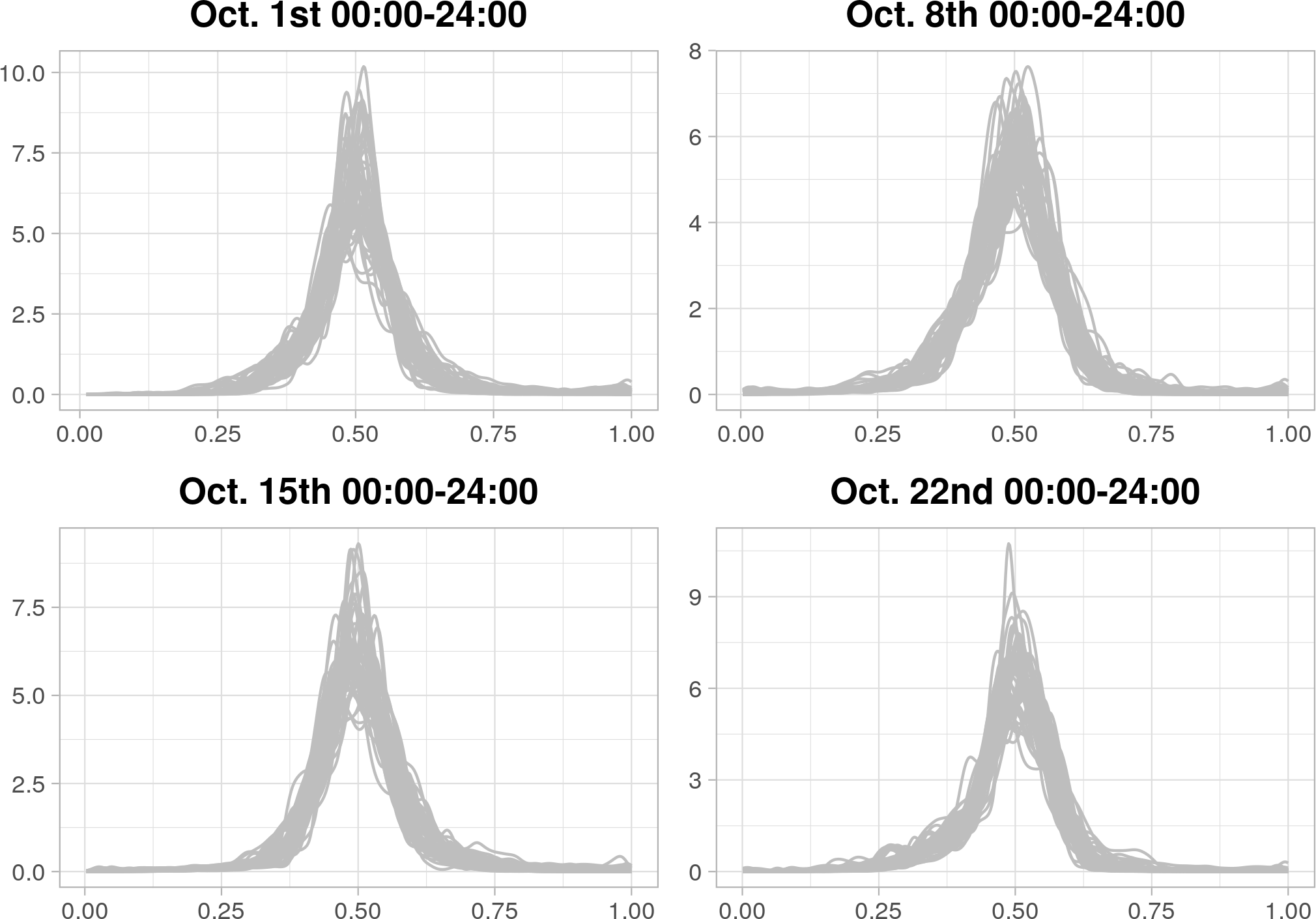}
\caption{Density curves representing fifty simulations of the (normalised) relative risk function considering four days.}
\label{fig:relrisk-curves}
\end{figure}

We repeated the procedure listed in Algorithm~\ref{alg:alg1} fifty times obtaining, for each pixel composing the road network, several estimates of the relative risk function. The smoothed curves displayed in Figure~\ref{fig:relrisk-curves} represent the distribution of $\hat{\rho}(\bm{s})$ for each time period and for each simulation. In all cases, these curves are concentrated around 0.5 and the value of $\hat{\rho}(\bm{s})$ lies between 0.4 and 0.6 for approximately 80\% of all road pixels. Moreover, the supplementary material reports the results obtained when testing the spatial accuracy for different time periods and training sets. In particular, the procedure detailed above was replicated by considering two alternative time intervals of six and twelve hours, respectively. We also tested the sensitivity of our results by shifting the training and test sets ahead of one and two months, respectively. We found that the suggested approach successfully predicts future events in all scenarios under consideration.

Finally, we re-adapted the ideas in \textcite{kelsall1998spatial, kelsall1995kernel} to our context by constructing a procedure that investigates the spatial variation of $\rho(\bm{s})$ and tests whether two processes defined on a common network $L$ (e.g. observed and simulated future emergency interventions) have the same intensity function. More formally, given two point patterns $\bm{A}$ and $\bm{B}$ with, respectively, $n_{\bm{A}}$ and $n_{\bm{B}}$ observations, the aforementioned papers proposed a Monte Carlo test for departure from a null hypothesis of random labelling, i.e. $H_0: \tilde{\rho}(\bm{s}) = \log\lbrace g_{\bm{A}}(\bm{s}) / g_{\bm{B}}(\bm{s})\rbrace = 0$ or, equivalently, $H_0: \rho(\bm{s}) = \frac{g_{\bm{A}}(\bm{s})}{g_{\bm{A}}(\bm{s}) + g_{\bm{B}}(\bm{s})} = 0.5$. The test was implemented by generating $m$ new datasets which are consistent with $H_0$ but, otherwise, have similar characteristics with respect to the original processes. Moreover, the authors used the following test statistics
\begin{equation}
t_j = \int \log\left\lbrace\frac{\hat{g}_{j, \bm{A}}(\bm{s})}{\hat{g}_{j, \bm{B}}(\bm{s})}\right\rbrace^ 2 \,\text{d}(\bm{s}),\quad j = 1, \dots, m,
\label{eq:test-statistics}
\end{equation}
where $\hat{g}_{j, \bm{A}}(\bm{s})$ and $\hat{g}_{j, \bm{B}}(\bm{s})$ respectively represent kernel estimates of the density function of the two processes $\bm{A}$ and $\bm{B}$ for the $j$th simulated dataset and $t_O$ is the observed value of the test statistics on the original point patterns. The p-value can be computed as $p = \frac{k + 1}{m + 1}$, where $k$ is the number of times that $t_j > t_O$.

Conditional on the location of the points, the null hypothesis states that the probability that a given event belongs to $\bm{A}$ instead of $\bm{B}$ does not depend on the spatial location and is constant over the region. Therefore, the generation of datasets under $H_0$ can be performed by combining the two original point patterns into a unique object and randomly labelling $n_{\bm{A}}$ of them as coming from process $\bm{A}$ and the remaining ones as type $\bm{B}$.

The algorithm described in the previous paragraphs was adjusted for the comparison of observed and simulated EMS data on a linear network as performed in this paper by adopting the following two modifications: 
\begin{enumerate}[noitemsep]
\item the integral in Equation~\eqref{eq:test-statistics} is computed over the network $L$ with arc-length measure and the density estimates are derived applying the weighted kernel approach described in Section~\ref{sec:spat-mod};
\item considering the intrinsic variation of simulated future ambulance interventions, the Monte Carlo test was repeated $\tilde{n}$ times and, for each simulation, the p-value was derived generating $\tilde{m}$ datasets under the null hypothesis of random labelling using the technique described before.   
\end{enumerate}

\begin{figure}
\centering
\includegraphics[width=0.7\linewidth,draft=false]{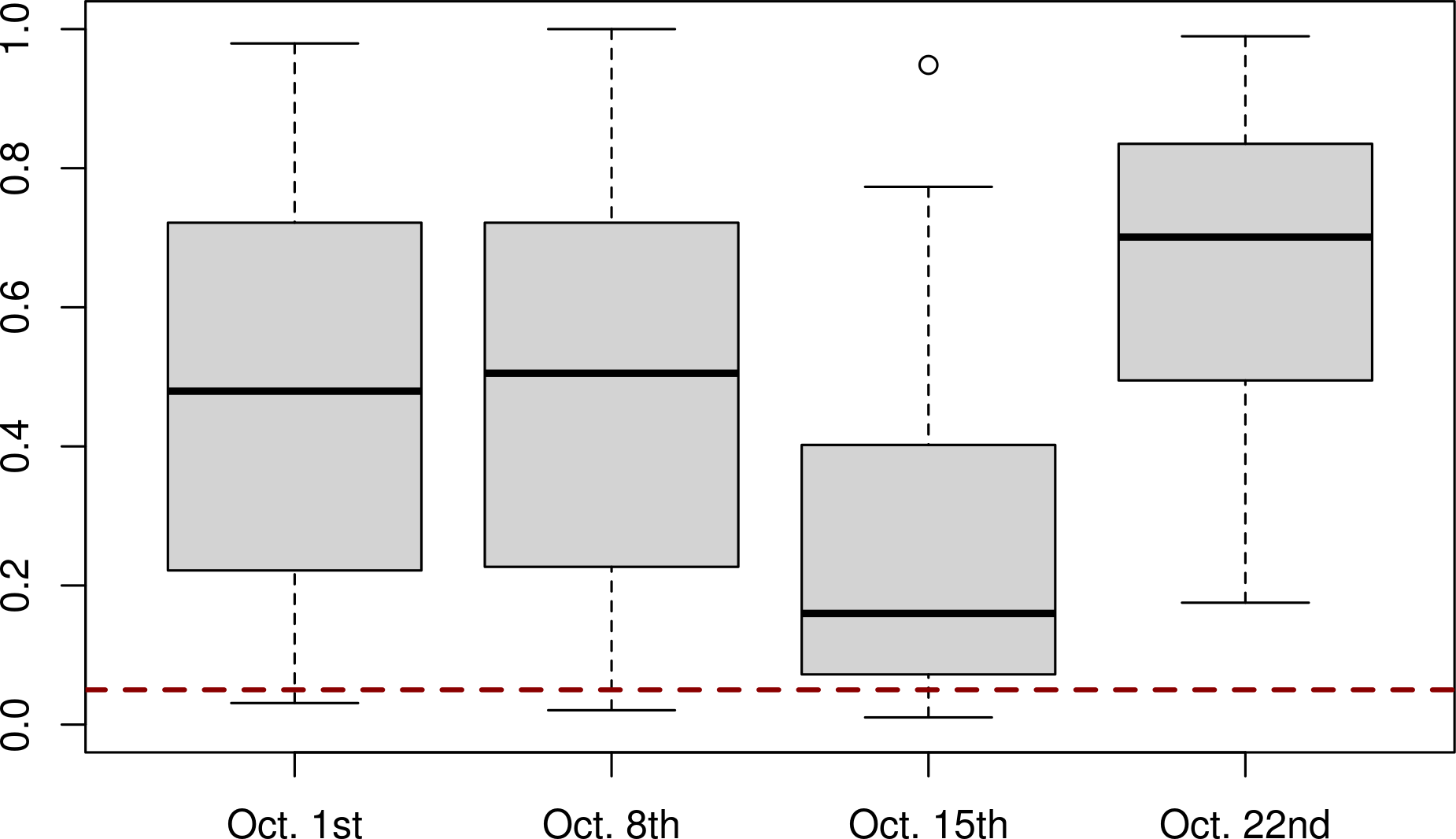}
\caption{Boxplots displaying the pvalues of Monte Carlo tests exploring the spatial variation of the relative risk function that compares observed and predicted EMS events. The dashed red line denotes the $\alpha = 0.05$ level.}
\label{fig:boxplot-pvalues}
\end{figure}

The results are reported in Figure~\ref{fig:boxplot-pvalues}. The four boxplots summarise the p-values obtained by comparing observed and predicted EMS data considering the same four time periods as before. For each time period, we generated $\tilde{n} = 64$ possible future scenarios using the procedure described in Algorithm~\ref{alg:alg1} and, for each scenario, we simulated $\tilde{m} = 96$ datasets under the random labelling hypothesis. We can clearly notice that, in all four cases, the relative risk function cannot distinguish between observed and predicted future interventions and we cannot reject the null hypothesis of random labelling. Therefore, this test highlights that the proposed model can successfully predict the future distribution of emergency data. 

\section{Additional comparisons with separable and planar models and further considerations}
\label{sec:compare}

As mentioned in Section~\ref{sec:intro}, this paper represents the first attempt to model ambulance interventions on a linear network considering an NHPP with a non-separable first-order intensity function. Although the proposed approach was found to perform reasonably well, a few aspects deserve further consideration. 
First, some comparisons to the previously proposed methods should be considered in order to appreciate how the conceptual improvement provided by our methodology translates into a practical improvement in a real-world application.
Second, although our model efficiently deals with the considered network (which includes the most important roads of Milan), we should also test whether this methodology effectively scales to larger networks such as the complete road network of cities like Milan which can be composed by hundreds of thousands of road segments. 
Third, some evaluations are in order regarding how much of a difference the methodological improvements of the proposed approach make for the ultimate application and how they translate into practical interventions on ambulance dispatch policy.
These three points are discussed in the rest of this section.

\subsection{Comparison with separable and planar models}

Sections~\ref{sec:separable-vs-non-separable} and~\ref{sec:planar-network-comparison} highlight the importance of the model extensions considered in this paper, namely modelling the spatio-temporal intensity in a non-separable manner while explicitly accounting for the network structure of the spatial domain. Hereinafter, the suggested methodology is compared to two different approaches adopted in previous papers dealing with EMS data, namely a) assuming separability of the first-order intensity function and b) ignoring the network structure of the road system and using a planar spatial support instead. 

\subsubsection{Separable first-order network-based intensity function}
\label{sec:separable-vs-non-separable}

\begin{figure}
\subfloat{\includegraphics[width=0.32\linewidth, draft=false]{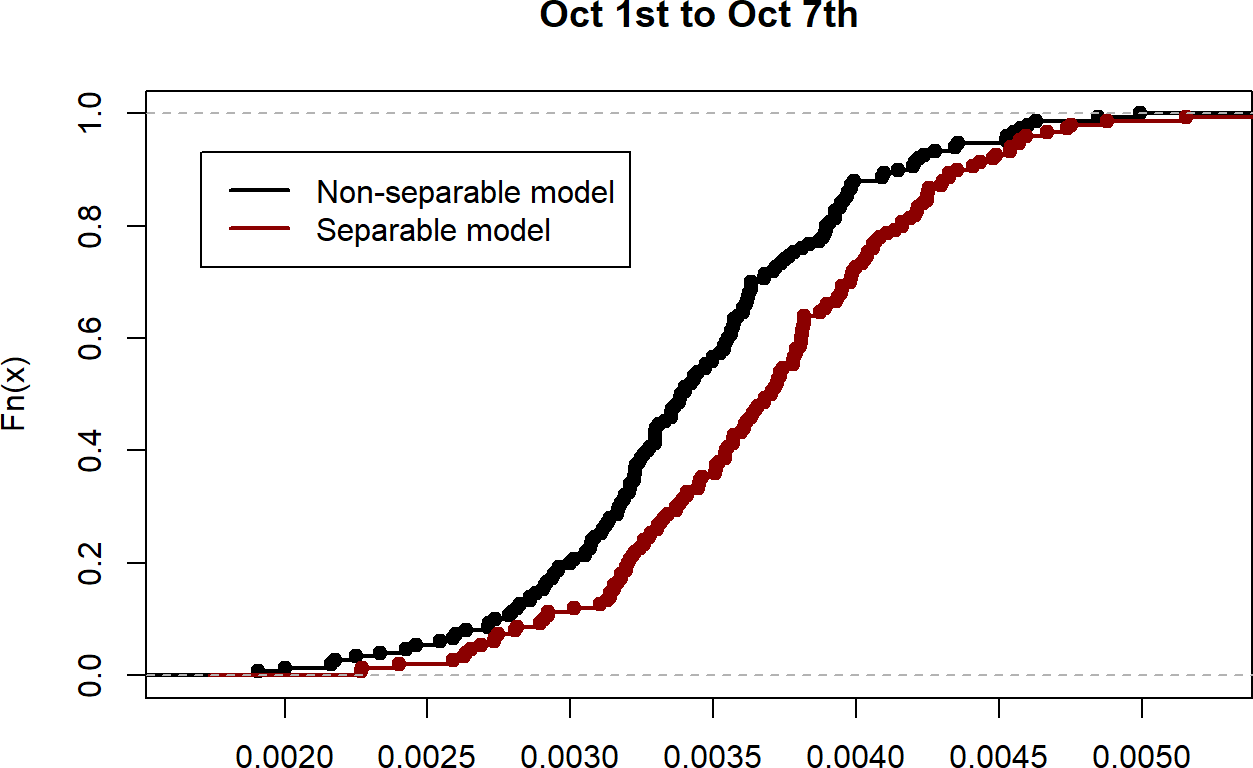}}
\hspace{0.01\linewidth}
\subfloat{\includegraphics[width=0.32\linewidth, draft=false]{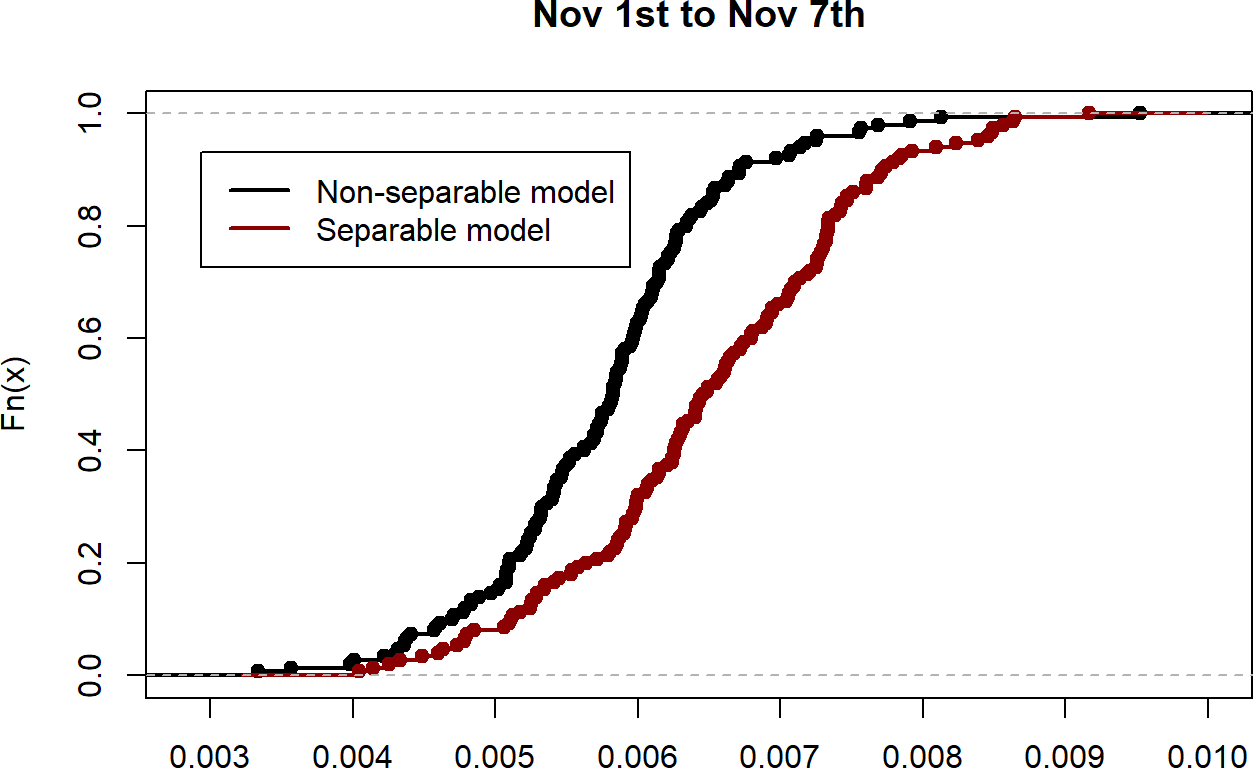}}
\hspace{0.01\linewidth}
\subfloat{\includegraphics[width=0.33\linewidth, draft=false]{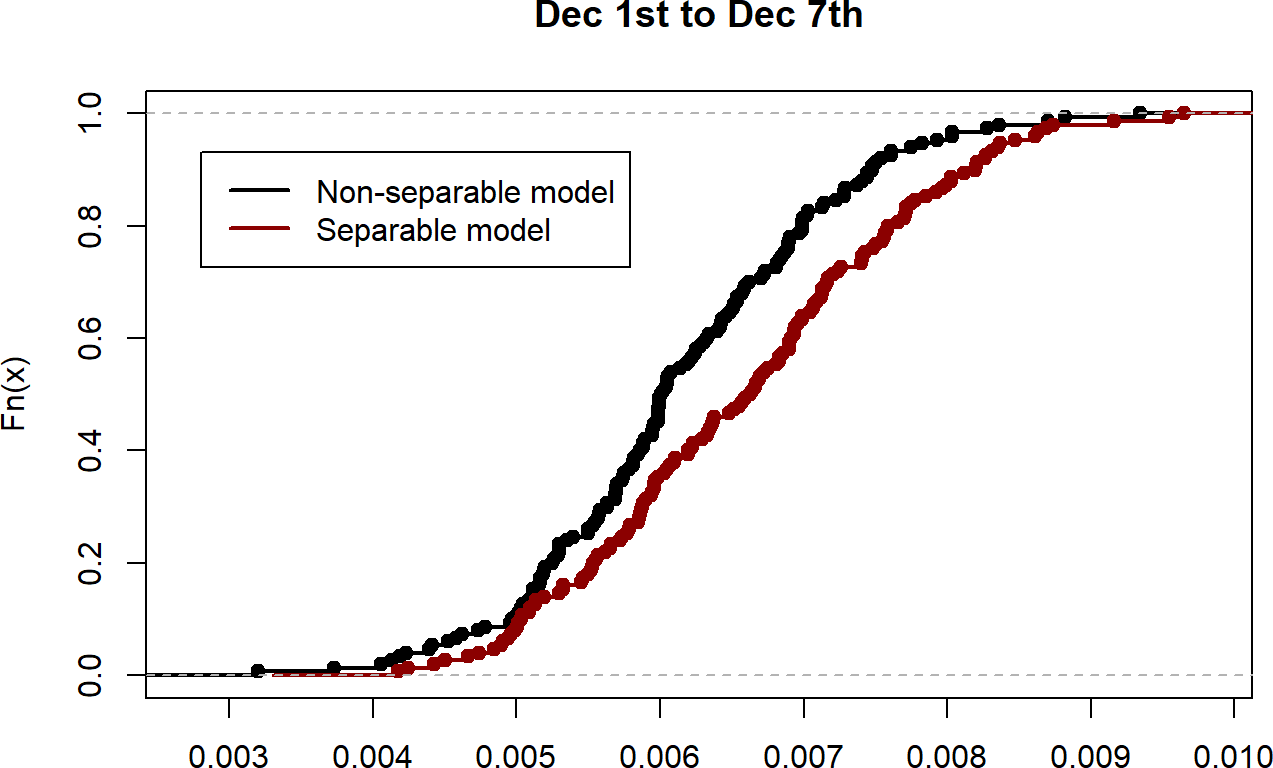}}
\caption{Comparison of separable and non-separable approaches for three different time periods. The three figures represent the ECDF of ISE obtained using 150 simulations.}
\label{fig:ISE-7d-test-separability}
\end{figure}

The impact of the non-separability assumption was tested by comparing our proposal to a simpler model that assumes a separable first-order intensity function and estimates the spatial dimension of the process via Equation~\eqref{eq:spat1b} assigning a unitary weight to each past observation. To sample from the simpler model we implemented a strategy analogous to the one described in the previous section and we compared the discrepancies between observed and predicted events at different levels of temporal aggregation for both models. More precisely, the two models were trained using the events that occurred before 2017-10-01 at 00:00 and we predicted $y_u$ observations for each hour $u$ of a  set $\mathcal{U}$ of future time periods as described in Algorithm~\ref{alg:alg1}. Then, the predicted and observed point patterns were aggregated in the considered temporal period and the predictive performances of the two strategies were evaluated by computing the Integrated Squared Error (ISE) defined by 
\begin{equation}
	\text{ISE} = \int_L \left(\hat{\eta}_{\text{pred}}(\bm{s}) - \hat{\eta}_{\text{obs}}(\bm{s})\right)^2 \ \text{d}_1(\bm{s}),
	\label{eq:ISE-separable}
\end{equation}
where $\hat{\eta}_{\text{pred}}(\bm{s})$ and $\hat{\eta}_{\text{obs}}(\bm{s})$ denote the smoothed spatial density at location $\bm{s} \in L$ obtained by the network convolution kernel \parencite{rakshit2019fast}. 

The procedure was repeated one hundred and fifty times, obtaining several estimates of the ISE for the separable and non-separable strategy. Figure~\ref{fig:ISE-7d-test-separability} reports the Empirical Cumulative Distribution Function (ECDF) of ISE criterion considering three time windows of 7 days located farther and farther from the end of the training set. The black curve denotes the non-separable model, whereas the red curve represents the separable one. We notice that, in all cases, the non-separable approach has superior predictive performances exhibiting lower ISE values and pointing out that the non-separability assumption plays a key role in the prediction of EMS events. 

In the supplementary material, we have reported the results obtained when applying the same procedure to temporal windows of 5 and 14 days. In all cases, the ECDFs show the same behaviour as in Figure~\ref{fig:ISE-7d-test-separability}, highlighting the stability of our findings. Furthermore, we compared each pair of curves using a Kolmogorov-Smirnov test and the non-separable model always outperformed the separable counterpart in all tested scenarios but one where the Kolmogorov-Smirnov was not significant. 

\subsubsection{Planar intensity function}
\label{sec:planar-network-comparison}

Hereinafter we evaluate the importance of taking the street network into account. To this end, we compared the approach suggested in this paper to another model developed using the same statistical structure but on a planar spatial domain (i.e. the polygon delimiting the city of Milan). 
Following the procedure detailed in Algorithm~\ref{alg:alg1} and adopting the same time windows considered in  Section~\ref{sec:separable-vs-non-separable}, we trained the two models and sampled $y_u$ points for each hour $u$ of the future time period $\mathcal{U}$. Finally, after aggregating the points, we compared predicted and observed interventions using the relative Integrated Squared Error (rISE) criterion that reads 
\begin{equation}
\text{rISE}_{\text{net}} = \int_L \left(\frac{\hat{\eta}_{\text{pred}}(\bm{s}) - \hat{\eta}_{\text{obs}}(\bm{s})}{\hat{\eta}_{\text{obs}}(\bm{s})}\right)^2 \ \text{d}_1(\bm{s}),
\label{eq:rISE-network}    
\end{equation}
in case of point patterns defined on network support and 
\begin{equation}
\text{rISE}_{\text{planar}} = \int_W \left(\frac{\hat{\eta}_{\text{pred}}(\bm{s}) - \hat{\eta}_{\text{obs}}(\bm{s})}{\hat{\eta}_{\text{obs}}(\bm{s})}\right)^2 \ \text{d}(\bm{s}). 
\label{eq:rISE-planar}    
\end{equation}
in case of a planar domain.

The quantities in Equation~\eqref{eq:rISE-network} have been introduced before, whereas the terms $\hat{\eta}_{\text{pred}}(\bm{s})$ and $\hat{\eta}_{\text{obs}}(\bm{s})$ in Equation~\eqref{eq:rISE-planar} respectively denote the planar smoothed spatial density at location $\bm{s} \in W$ (where $W$ denotes the 2-dimensional domain) for predicted and observed interventions obtained via a classical planar kernel with Jones-Diggle's edge correction \parencite{jones1993simple}. We adopted the relative ISE criterion to compare the network and planar estimates since the two processes are defined on incompatible spatial domains, implying that the corresponding density functions have different orders of magnitude and making a direct comparison unfeasible. More precisely, the intensity function on a linear network has dimensions $1/m$ while its planar counterpart has dimensions $1/m^2$. Therefore, the ISE criterion as defined in Equation (11) would compare two quantities having dimensions $1/m^2$ and $1/m^4$, respectively. On the other hand, as we can see from Equations~(12) and~(13), the rISE criterion includes an additional term at the denominator of the two Equations that creates a unit-less ratio of intensities and removes the effects due to the different natures of the corresponding spatial domains.

\begin{table}
\centering
\caption{Numerical summary of the comparisons between network and planar approaches using the rISE criterion defined in Equations~\eqref{eq:rISE-network} and~\eqref{eq:rISE-planar} considering 3 time-windows of 7 days.}
\label{tab:rISE-planar-network}
\begin{tabular}{lc*{4}{S[
	round-mode = figures,
	scientific-notation = true
	]}}
\toprule Time window & Type & {Mean} & {Std. Dev.} & {0.25 Quantile} & {0.75 Quantile}\\
\midrule 
\multirow{2}{*}{Oct. 1st to Oct. 7th} 
& Network & 101287 & 39716.2 & 72664.37 & 122260.13 \\
& Planar & 17119821 & 11857262.4 & 10593933 & 18797738 \\
\multirow{2}{*}{Nov. 1st to Nov. 7th} 
		& Network & 209060 & 63040.82 & 167185.3 & 242637.2 \\
		& Planar & 29968828 & 14313215.81 & 20843099 & 34892442 \\
		\multirow{2}{*}{Dec. 1st to Dec. 7th} 
		& Network & 192979.8 & 123591.8 & 118756.3 & 216693.4 \\
		& Planar & 23940243.5 & 11225069.1 & 16863382 & 27310067 \\
		\bottomrule
	\end{tabular}
\end{table}

We repeated the procedure described above one hundred and fifty times obtaining several values of the rISE index for each time window. The results are summarised in Table~\ref{tab:rISE-planar-network}. We can clearly notice that the average rISE for the network approach is several times smaller than its planar counterpart, highlighting that the analysis of EMS interventions always requires appropriate considerations regarding the spatial support of the events. The same procedure was repeated for different time intervals of 5 and 14 days, obtaining similar conclusions. The results are summarised in the supplementary materials. 

\subsection{Scalability}
As already mentioned in Section \ref{sec:data}, the analyses reported in this paper are based on a subset of Milan's street network that includes the most important road types since the majority of ambulance interventions were geo-referenced on their proximity. This may rise some concerns about the scalability of the proposed methodology, i.e. the ability of our procedure to maintain effectiveness when applied to a larger network if this would be necessary under different circumstances. In the supplementary material, we summarise the results obtained when applying the statistical model detailed in Section \ref{sec:methods} to a larger spatial network composed by 118720 segments covering 4636km. It was created considering all road segments located in Milan that are available from OSM servers. These additional analyses allow us to assess the robustness to different spatial networks of the approach proposed in this paper for EMS data modelling and prove the excellent scalability of the suggested methodology with large spatial domains. More precisely, after downloading the network data from OSM servers, it took approximately 17 minutes to estimate the temporal model, the weight function and the KDE in Equation \eqref{eq:spat1b} on the extended network. Although the extended network is more than two times longer than the original one, the computational time required to perform the statistical analysis was definitely reasonable as compared to the time requested, 10 minutes in total, to analyse the data on the restricted network considered in the previous sections of this paper. In particular, thanks to the Fast Fourier Transform algorithm adopted in the kernel estimator, fitting the statistical model on the two networks requires roughly the same computational effort. Further details on these comparisons are reported in the supplementary material.

\subsection{Use for policy interventions}
Ambulance dispatch planning requires careful consideration of a number of different factors and the model proposed in this paper can support this activity by providing an estimate of the potential demand for interventions in different parts of the city network.

\begin{figure}[tb]
	\centering
	\includegraphics[width=0.85\linewidth]{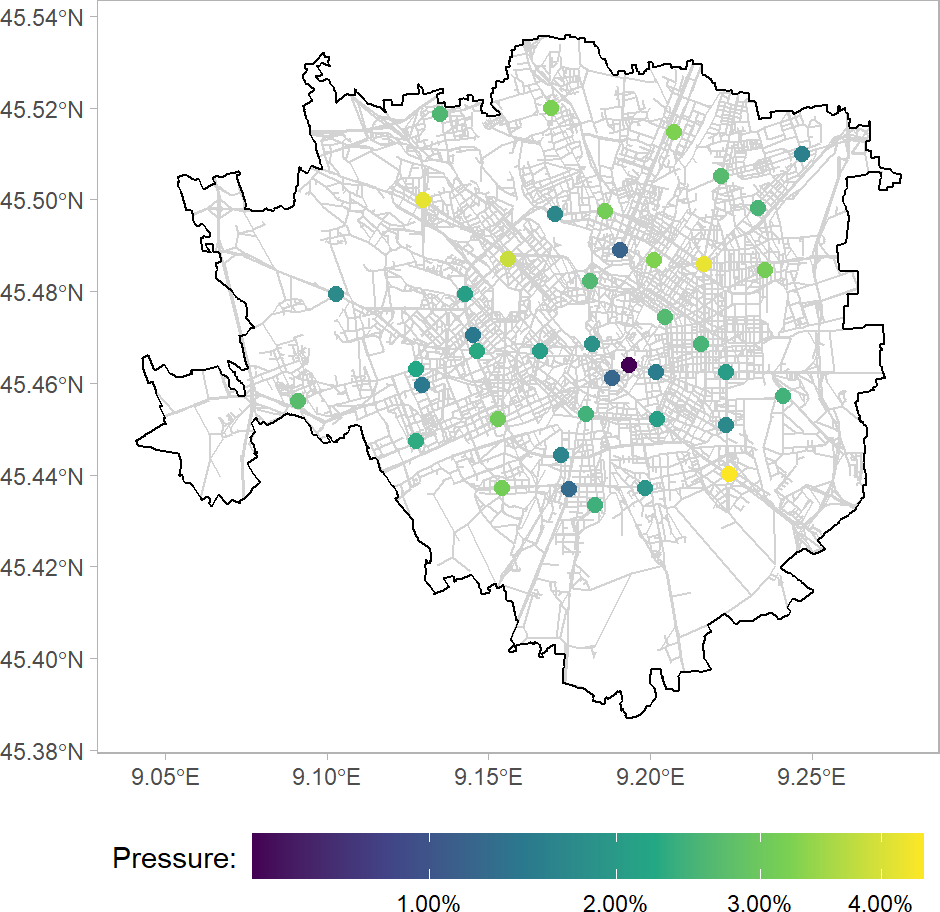}
	\caption{Predicted pressure on the 42 ambulance stations of Milan for January 3, 2018.}
	\label{fig:pressure}
\end{figure}

Hereinafter we have considered how the analysis described in the previous sections may help local EMS agencies to manage the ambulance rescue points. In fact, in the city of Milan, there are 42 locations (such as dedicated squares or parking spots) where the ambulances can park during the day or the night while waiting for a request of intervention. As shown in Figure~\ref{fig:pressure}, these rescue points are placed in strategic areas of the city network identified to optimise access time to patients and provide good coverage of the territory of the city.

Using the model described in the previous sections, we are able to anticipate the pressure that each of these stations would suffer in terms of requests for intervention on a given day. We exemplified this point using the data collected between 2015-01-01 and 2017-12-31. We fitted our model to the data and used it to simulate the locations of the ambulance interventions a few days ahead. We then calculated the minimum distance on the road network between each simulated event and the closest ambulance station, assuming that this station would be the one to be activated in order to provide the most efficient reaction. In this way, we were in the position to estimate the potential pressure on each station that is expected on the considered day, where the potential pressure is calculated as the percentage of interventions that are closer to this station than to any other station in the city. In order to compensate for the simulation variability, we simulated 150 point patterns using the fitted model and averaged the percentage over the performed simulations. 

This information may allow the authorities to allocate some extra ambulance crews, if available, to those stations that are expected to be exposed to high pressure or move ambulances stationed on low-stress points there. 
Given the low computational time (the previous simulations take only a few minutes) this analysis can be conducted on a daily basis to maintain an up-to-date intervention units system or, at least, to have a benchmark to compare the spatial allocation of ambulance crews. 

In addition, the model developed in this paper can be potentially adopted to evaluate how efficient ambulance dispatches have been on a given day by comparing, retrospectively, the actual pressure observed in the considered day to the one estimated using the approach described above. We performed this analysis on October 1, 2017, finding only a mild agreement between the two quantities: only 8 of the top 15 stressed stations were in the top 15 predicted pressures. Although several other factors may impact on the observed pressure of a given station in a certain day  (e.g. rerouting of ambulances from/to other parts of the city or temporary traffic jams occurring nearby the ambulance point or the location of intervention), discrepancies between potential and actual pressure may suggest a rethinking of the criteria adopted for ambulance dispatches in order to speed up the service and improve its efficiency.

\section{Conclusions}
\label{sec:conclusions}

In this paper, we investigated the spatio-temporal distribution of approximately $480,000$ ambulance interventions that occurred in the city of Milan from 2015-01-01 to 2017-12-31. Unlike several previous approaches, we assumed that the emergency events represent a realisation of a spatio-temporal point process occurring on a road network, i.e. a geo-located graph structure representing road segments and street junctions. 

A preliminary exploratory analysis, summarised in Section \ref{sec:data}, revealed that the temporal evolution of the events presents several types of seasonalities due to hourly, daily and weekly patterns. We also observed the presence of space-time interactions in the hourly distribution of the events, which motivated the adoption of a non-separable statistical model. More precisely, after dividing the interventions into one-hour intervals, we assumed that, for each time period, the ambulance dispatches represented a realisation of an NHPP on a linear network with a non-separable first-order intensity function. The temporal component was modelled via a semi-parametric Poisson regression with deterministic temporal covariates. Considering the results of the exploratory analysis, the annual patterns were included with a linear term, the hourly and weekly trends were smoothed using cyclic cubic regression splines, whereas the daily effects are included using dummy variables. 
The spatio-temporal component was modelled by a weighted kernel estimator. The weights were used to capture the space-time interactions of EMS data, trying to grasp the temporal regularities in the emergency interventions and induce non-separability into the spatio-temporal intensity.

We found that the temporal Poisson model fits the EMS counts well and the deterministic temporal components successfully approximate the hourly, daily, and weekly patterns. The weight function also adequately mirrors the temporal seasonalities displayed by the ACF of EMS counts. The spatial and spatio-temporal dynamics were exemplified 
considering 
two future time periods: 2018-01-03 at 03:00 and 2018-01-03 at 15:00. Our results highlight that ambulance interventions are more spread in the municipality during the night, whereas they tend to cluster in the city centre during working hours. In both cases, the main train station, a few popular squares, and a retirement house stand out. 

The predictive accuracy of our proposal was tested using the relative risk function by comparing observed and predicted ambulance interventions for four different days. In all cases, the relative risk is concentrated around 0.5, implying that the model successfully predicts future events. A series of Monte Carlo tests confirmed that conclusion.

Finally, we demonstrated that the approach proposed in this paper improves over the methodologies previously adopted for modelling EMS data, taking into account both the network structure of the spatial domain and the non-separability of the spatio-temporal intensity function. We also found that this approach scales well to very large networks, hence it proves to be particularly suitable to manage real-world applications.

To conclude, we remark that the main challenges in this paper stem from the spatio-temporal dynamics and the specific spatial support of our data. First, the exploratory analysis suggested an interaction in the spatial and temporal components, requiring a non-separable structure when modelling the ambulance intervention process. Second, the spatial nature of the data also suggested that linear networks are the most appropriate spatial domain for modelling EMS data. Third, the geographic region represents a large metropolitan area and the huge number of interventions required the adoption of fast statistical techniques. This latter point may also imply that the spatial and temporal variability can be impacted by secondary variables possibly measured both at the areal level (e.g. population density) and at the network level (e.g. road types, traffic flows, commuting patterns or other variables representing specific anthropic activities at a given point of the network). However, the main purpose of this paper is the spatial and short-term temporal prediction of ambulance interventions and, according to our experience, regionalised time-varying covariates are difficult to obtain at the desired spatial and temporal levels and their inclusion is scarcely impactful. Nevertheless, in future works, it might be desirable to develop parametric or semi-parametric models that allow the introduction of such explanatory spatial covariates in the intensity function. Furthermore, it should be pointed out that some of the aforementioned variables are typically recorded only at the areal level (e.g. census wards or traffic zones) and their inclusion in a network model presents several layers of complexity. In fact, the projection of areal data into a linear network may induce abrupt changes in the covariate (every time a segment intersects different areas) or imprecise measurements, hence requiring further modelling care.

We point out that, considering the complexities detailed before, machine learning (ML) methods (such as classification trees or neural networks) may represent a promising approach to analyse spatial and spatio-temporal point patterns. However, to the best of our knowledge, the literature is extremely scarce in this field and only a few recent papers exist addressing this aspect. For example, \textcite{yuan2019variational} merge the theory of classical kernel density estimation with variational autoencoders to develop a model for the analysis of spatial inhomogeneous Poisson processes. \textcite{mateu2022spatial} provide a mathematical framework for coupling neural network models with the statistical analysis of planar point patterns focusing on point processes with multiple groups observed for $T \ge 2$ times, whereas \textcite{jalilian2022assessing} develop a Siamese Neural Network Discriminant Model to evaluate the similarities between spatial point patterns obtaining superior performances than the classical statistical tools (i.e. the $K$ function). However, it should be noted that the deep learning methods introduced in the aforementioned papers typically require that the point pattern is reduced to a two-dimensional grid of cell counts that is treated as an image, i.e. a set of pixel values; hence planar spatial support is assumed for the data. The adaptation of those techniques to the analysis of linear network data requires substantial methodological improvements which are beyond the scope of this paper. 

A further extension is to move towards statistical models that account for data clustering following different routes, for instance, a double stochastic process, such as the Cox process where a stochastic component is included in the intensity function to deal with the unexplained space-time variation. A natural solution, in this case, would be to adopt an inhomogeneous log-Gaussian Cox process \parencite{moller1998log}, already proposed for modelling ambulance interventions by \textcite{bayisa2020large}. However, moving to the latter approach requires a substantial amount of methodological development since defining a proper covariance function for the stochastic component of the intensity function on a linear network spatial support is not straightforward.

\section*{Acknowledgements}

Map data copyrighted OpenStreetMap contributors and available from \url{https://www.openstreetmap.org.} We greatly acknowledge Doct. Piero Brambilla, Doct. Andrea Pagliosa and Doct. Rodolfo Bonora, the regional experts of ambulance interventions working for the local EMS (AREU: Azienda Regionale Emergenza Urgenza). They provided us the ambulance intervention data used in this paper and, more importantly, priceless and continuous assistance to properly understand the complexity of an EMS system and the peculiarities of this problem. We also greatly acknowledge the DEMS Data Science Lab for supporting this work by providing computational resources. J. Mateu is partially supported by grant PID2019-107392RB-I00 from the Spanish Ministry of Science.

\end{refsection}
\newrefcontext[sorting=nty]
\printbibliography[section=1]
\newrefcontext[sorting=ynt]

\clearpage 

\appendix

\begin{refsection}

\section{Supplementary Material}

\subsection{Additional results regarding the spatial validation procedure}

\begin{figure}
	\centering
	\includegraphics[width=\linewidth]{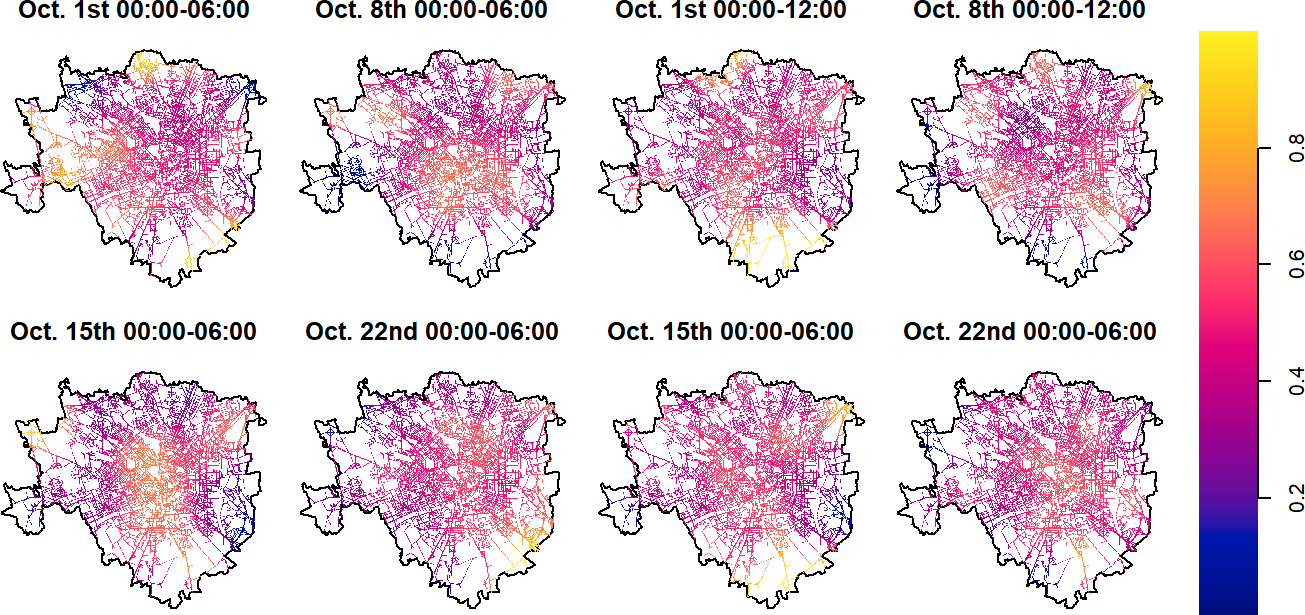}
	\caption{Maps of the (normalised) relative risk function for six and twelve hours intervals in four different days.}
	\label{fig:relrisk-maps-6h-12h}
\end{figure}

\begin{figure}
	\centering
	\includegraphics[width=\linewidth]{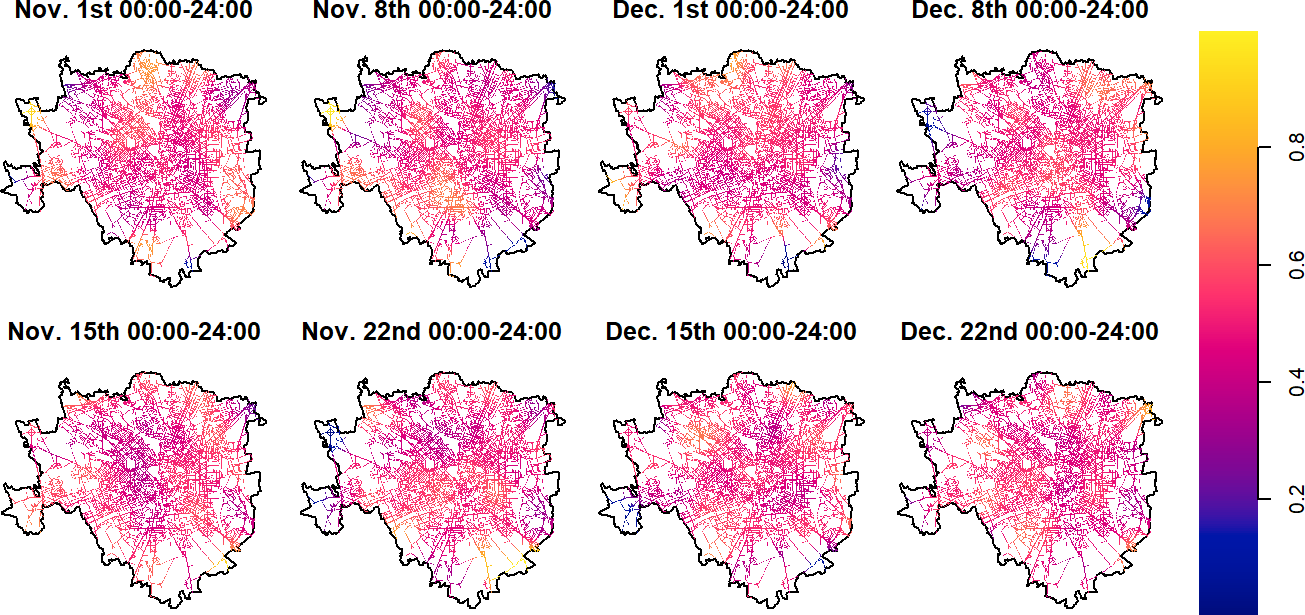}
	\caption{Spatial representations of the (normalised) relative risk function considering four days in November and December 2017. The weighted network KDE was trained considering all ambulance interventions until the end of October and November, respectively.}
	\label{fig:relrisk-maps-nov-dec}
\end{figure}

The procedures detailed in Section 5 of the paper were also run considering different time intervals and different training sets. The results are summarised hereinafter. Figure \ref{fig:relrisk-maps-6h-12h} considers eight time intervals of six and twelve hours, spread over the same days displayed in Figure 12 of the paper. Despite a slightly lower spatial accuracy due to the higher temporal granularity, the predicted EMS interventions approximate the observed events reasonably well. Approximately 65\% and 70\% of all road segments registered a value of $\hat\rho(\bm{s})$ between 0.4 and 0.6, respectively. Figure \ref{fig:relrisk-maps-nov-dec} depicts the (normalised) relative risk function estimated using a spatio-temporal model trained considering all EMS interventions that occurred until the end of October and November, respectively. We predicted the future emergency events considering the same days as in Figure 11 shifted by one and two months, respectively. In particular, 01-11, 08-11, 15-11, and 22-11 were considered in the first scenario, whereas 01-12, 08-12, 15-12, and 22-12 were analysed in the second scenario. The eight maps show that our results are also robust when predicting different months. In both scenarios, approximately 85\% of all road segments registered a value of $\hat\rho(\bm{s})$ between 0.4 and 0.6. 

\subsection{Comparing with the extended network}

As mentioned in the paper, the model detailed in Section 3 was also estimated considering a larger spatial network than the one presented in the manuscript. In particular, the extended network was created using OSM data for every type of street segment in the road network of Milan. It is composed by approximately 115,000 segments covering more than 4500km.

\begin{figure}
	\centering
	\subfloat[2018-01-03 03:00 \label{fig:pred-spat-dens-extended-03}]{
		\includegraphics[width=0.45\linewidth]{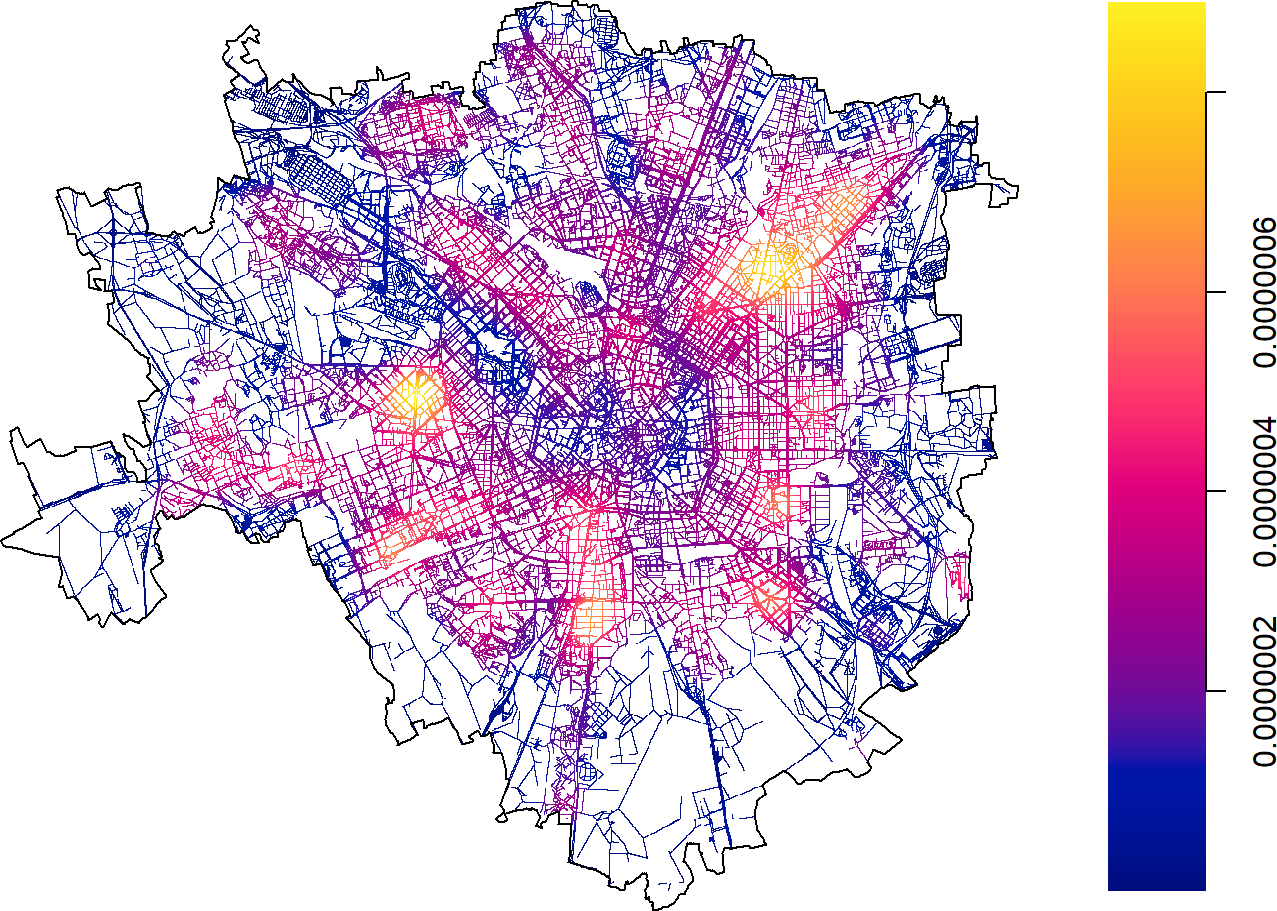}
	}
	\hspace*{0.25cm}
	\subfloat[2018-01-03 15:00 \label{fig:pred-spat-dens-extended-15}]{
		\includegraphics[width=0.45\linewidth]{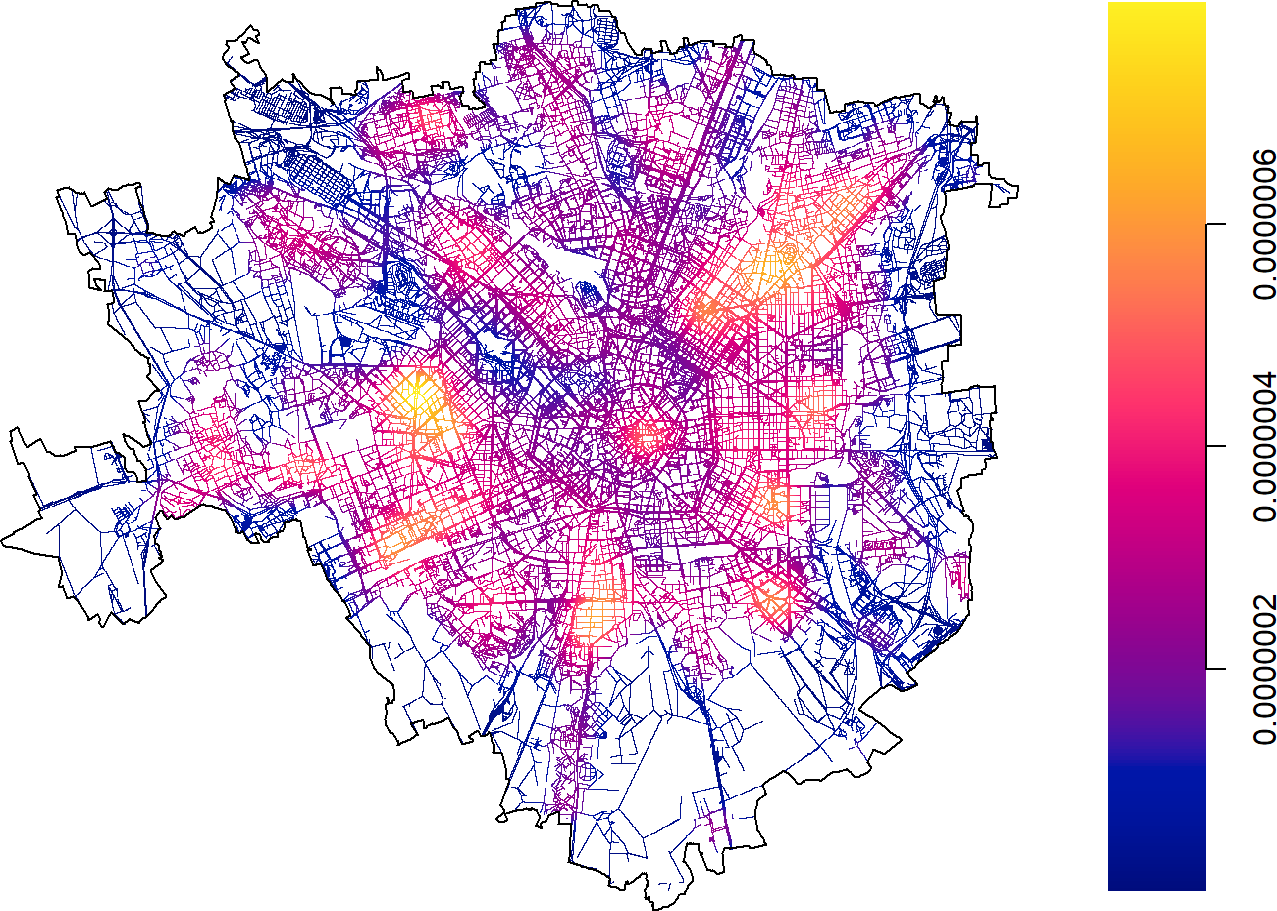}
	}
	\caption{Estimates of spatial density function $\hat{g}_{u}(\bm{s})$ considering a model estimated on the extended spatial network and two future time periods: 2018-01-03 at 03:00 (a) and 2018-01-03 at 15:00 (b).}
	\label{fig:pred-spat-dens-extended}
\end{figure}

Figure~\ref{fig:pred-spat-dens-extended} shows the results obtained when predicting the spatial density for the same time periods considered in the paper (i.e. 2018-01-03 at 03:00 and 2018-01-03 at 15:00). By comparing Figure~9 in the paper and Figure~\ref{fig:pred-spat-dens-extended}, we can first observe that the extended road network covers a larger area and is unevenly denser. This is particularly evident in several neighbourhoods located in the north-west of the city. However, hotspots were found in the same areas, i.e. the central station, an important square (Piazzale Corvetto), and several retirement castles. 
The two maps in Figure~\ref{fig:pred-spat-dens-extended} clearly represent the temporal evolution of the spatial patterns of EMS interventions. The map on the left was estimated considering a night time, hence it highlights night-life areas. On the other hand, the map on the right-hand side of the figure highlights the city centre and several working places near the Duomo of Milan, since it was estimated considering a specific hour in the afternoon.

Finally, as far as the scalability of the suggested methodology is concerned, Table~\ref{tab:computing-times} summarises the computing time for each step of the analysis using the two networks. In both cases, our model was estimated in a reasonable amount of time (10 minutes and 17 minutes, respectively). 
The first step involves the creation of the computational structure behind the linear network. This is the most demanding step of the analysis since converting OSM ways into a linear network with thousands of nodes and edges requires a non-trivial amount of computing power, that obviously increases with the size of the network \parencite{barthelemy2011spatial, pruvost2017exploring}. Although the extended network is definitely large, the computational burden remains feasible and the network can be built in a reasonable time. 
The temporal model and the weight function can be estimated really quickly and their computing time does not depend on the network size. 
Finally, as far as the implementation of the spatial KDE is concerned, we observe only a slight increase in the time necessary to calculate the estimator when the larger network is considered. In fact, the FFT algorithm adopted for KDE implementation \parencite{rakshit2019fast} permits the estimation of the spatial component of the process very quickly even for very large networks. According to our simulations, this is true also for even larger networks. The details of these simulations are not reported in this paper but are available upon request.

\begin{table}
	\caption{Table comparing the computing times for each step of the model described in Section~3 of the paper using the regular linear network (20064 segments covering 1949 km) and the extended linear network (99201 segments covering 4534 km). We can see that also in the former case the model can be estimated in a reasonable amount of time.}
	\label{tab:computing-times}
	\centering
	\begin{tabular}[t]{lrr}
		\toprule
		Step & Regular Network & Extended Network\\
		\midrule
		Build the road network & 5 minutes & 11 minutes\\
		Estimate the temporal model & 1.30 minute & 1.30 minute\\
		Estimate the weight function & 30 seconds & 30 seconds\\
		Estimate the spatial model (for each future time period) & 1.30 minute & 2.30 minutes\\
		Total & 10 minutes & 17 minutes\\
		\bottomrule
	\end{tabular}
\end{table}

\subsection{Additional results on the comparisons between separable/non-separable and network/planar approaches}

\begin{figure}
	\subfloat{\includegraphics[width=0.32\linewidth]{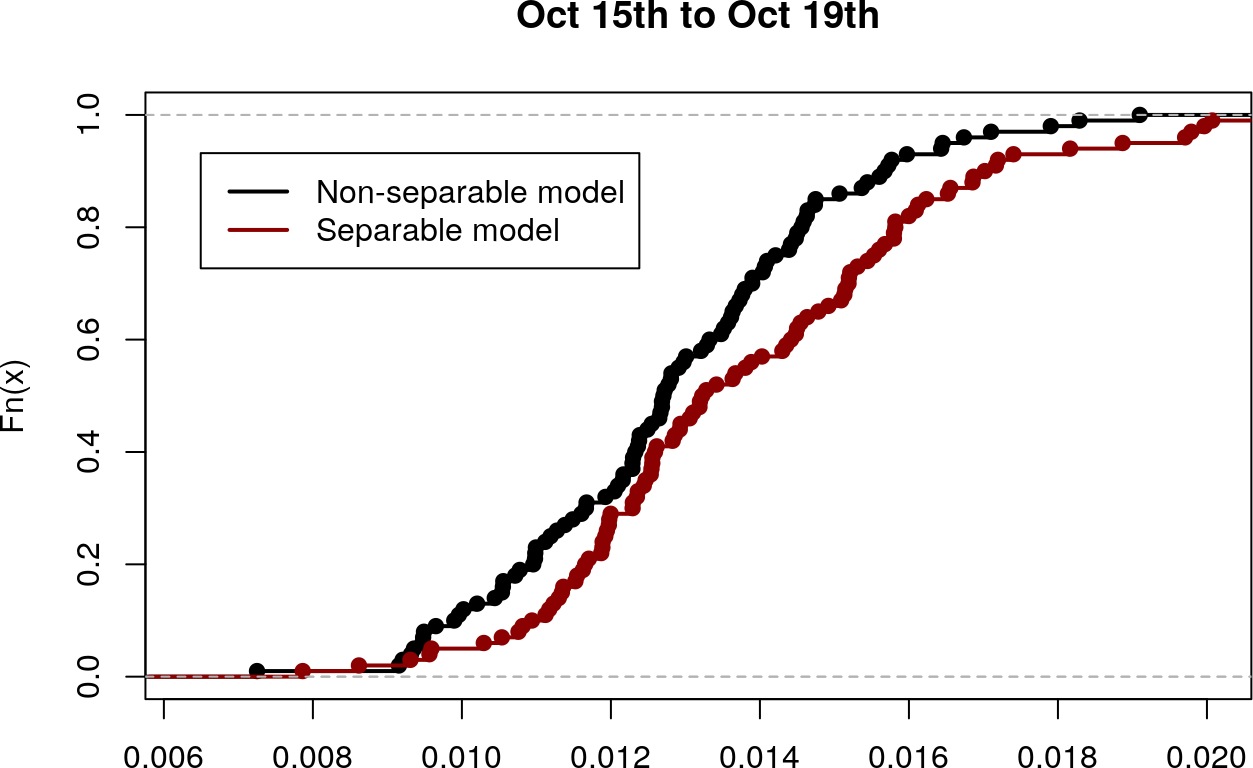}}
	\hspace{0.01\linewidth}
	\subfloat{\includegraphics[width=0.32\linewidth]{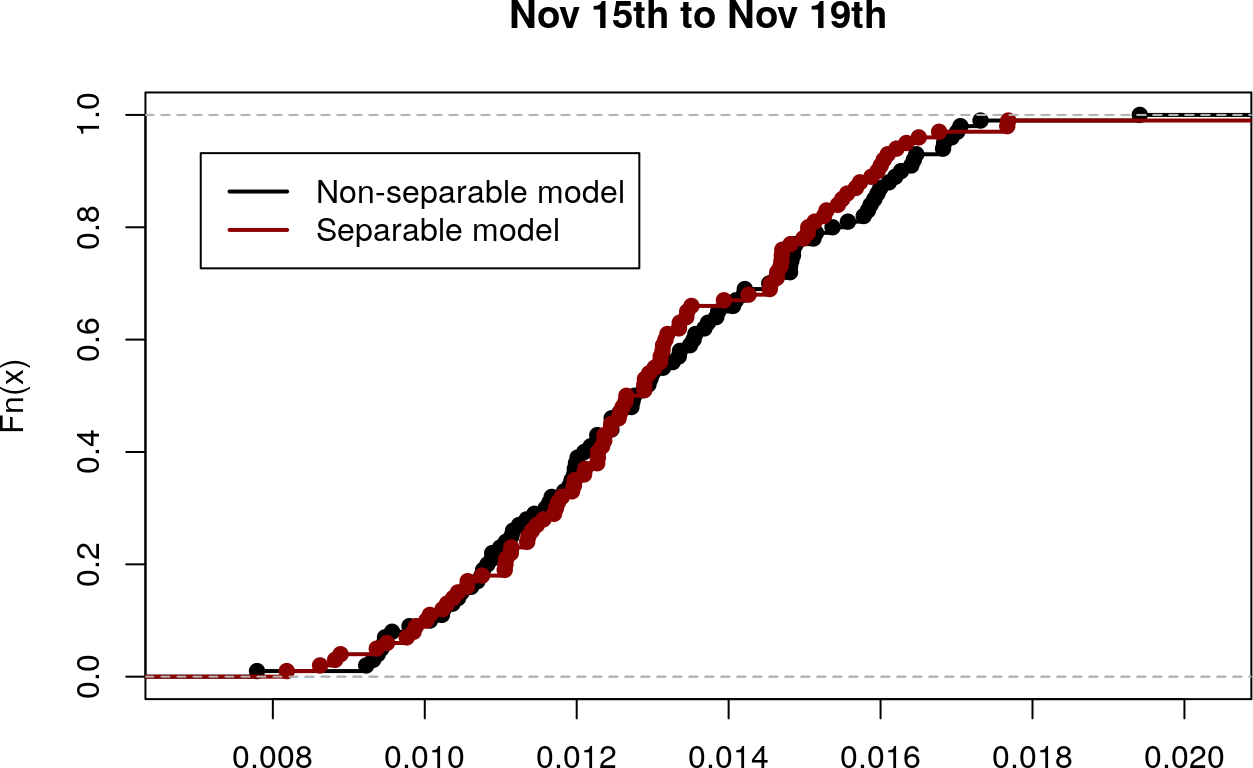}}
	\hspace{0.01\linewidth}
	\subfloat{\includegraphics[width=0.33\linewidth]{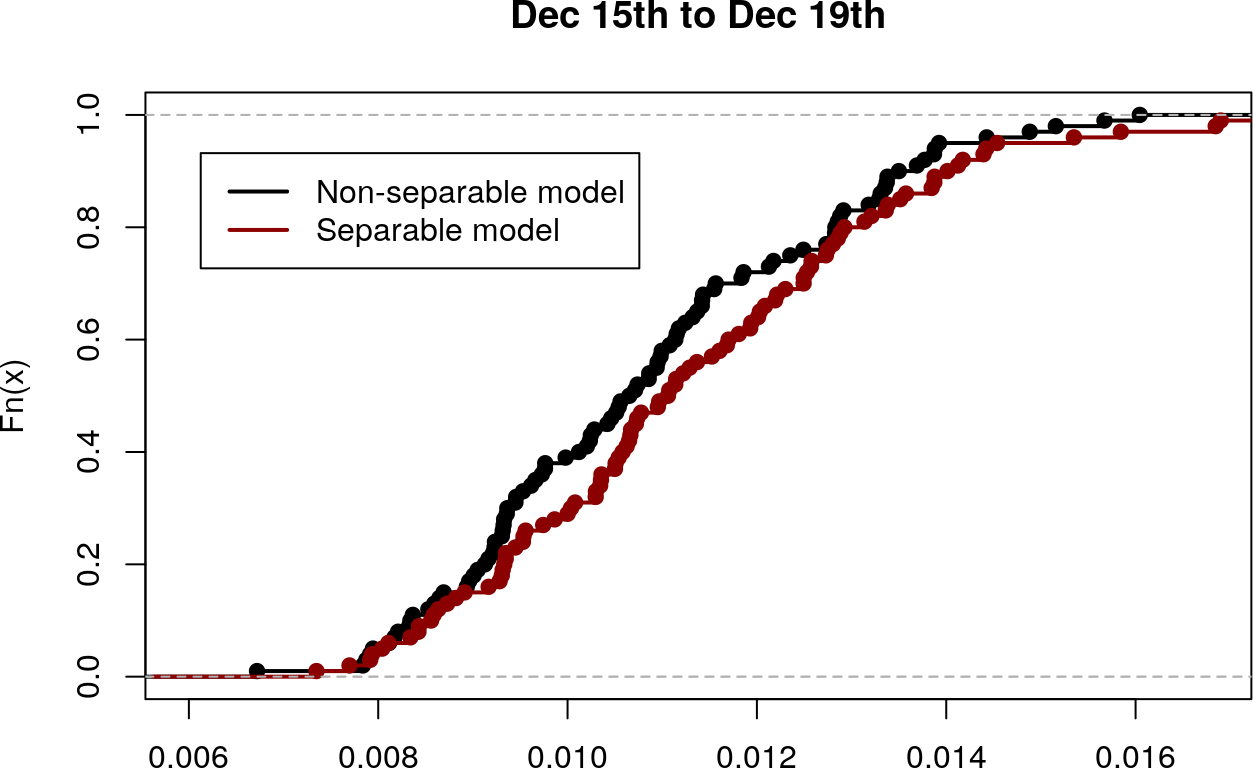}} \\
	\subfloat{\includegraphics[width=0.32\linewidth]{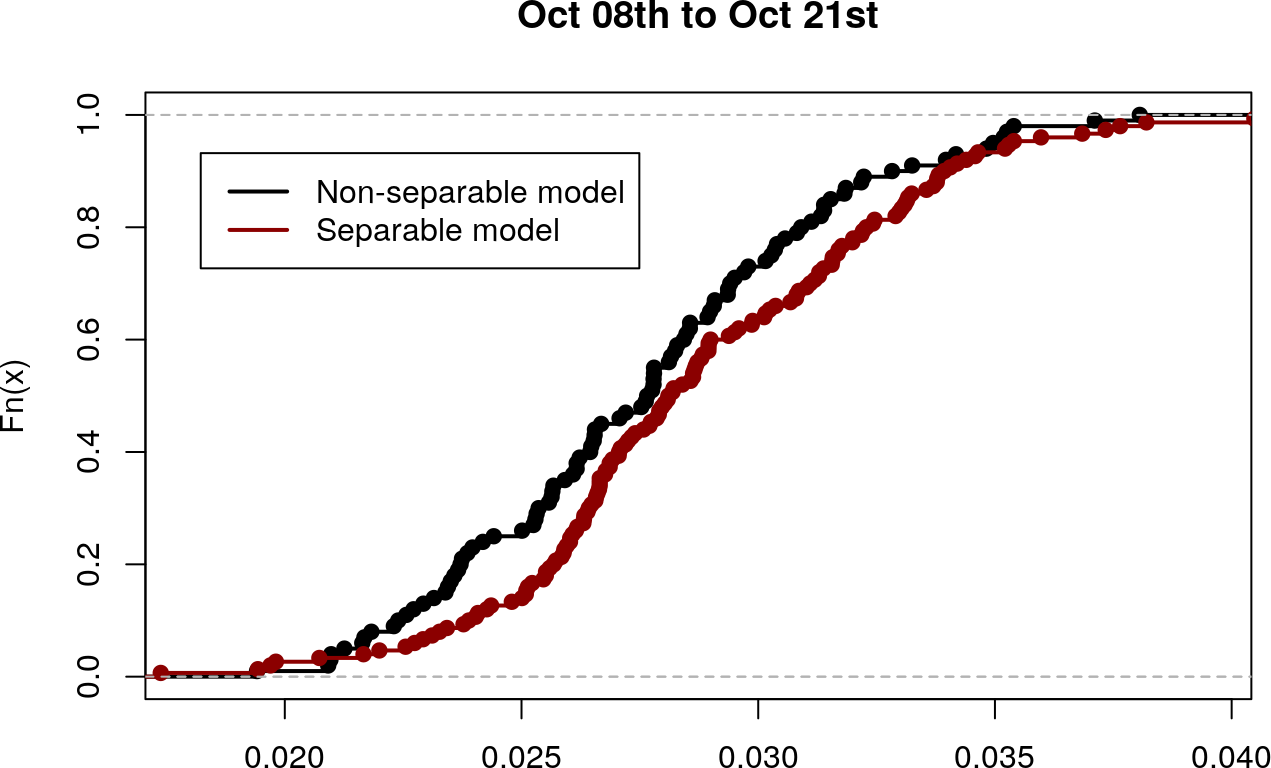}}
	\hspace{0.01\linewidth}
	\subfloat{\includegraphics[width=0.32\linewidth]{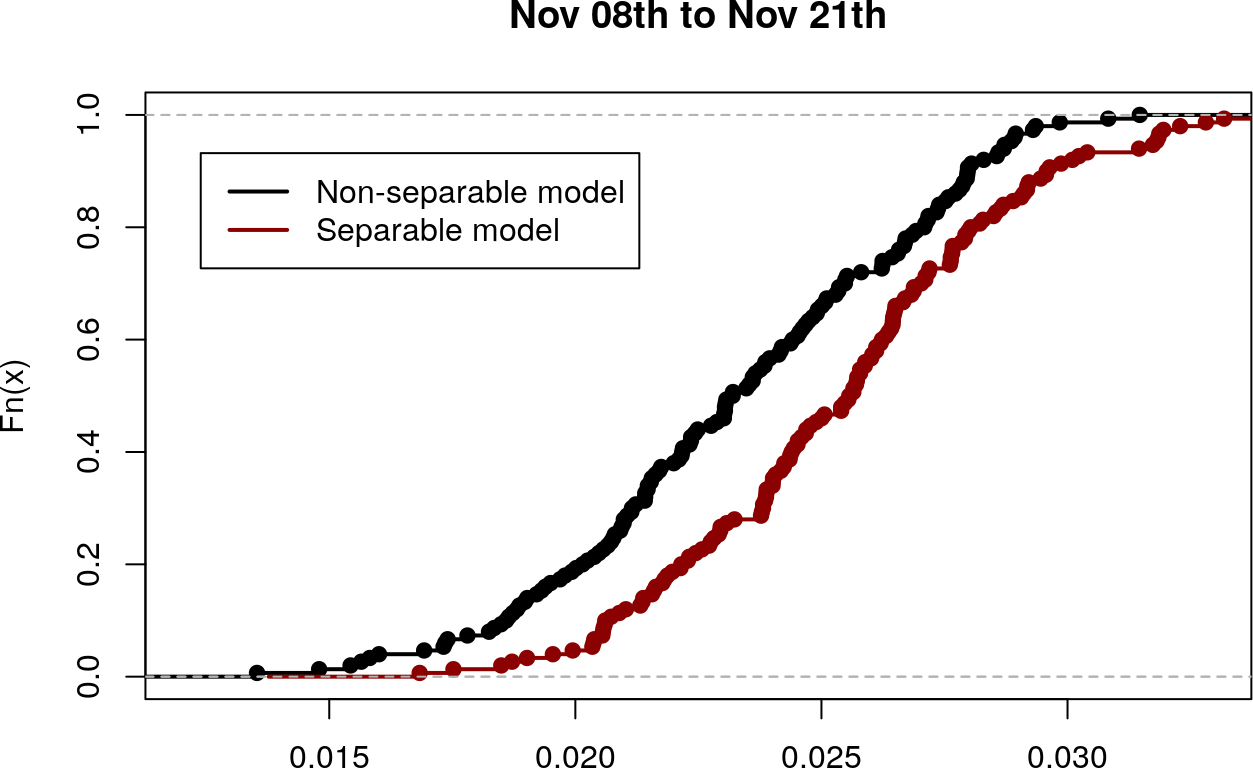}}
	\hspace{0.01\linewidth}
	\subfloat{\includegraphics[width=0.33\linewidth]{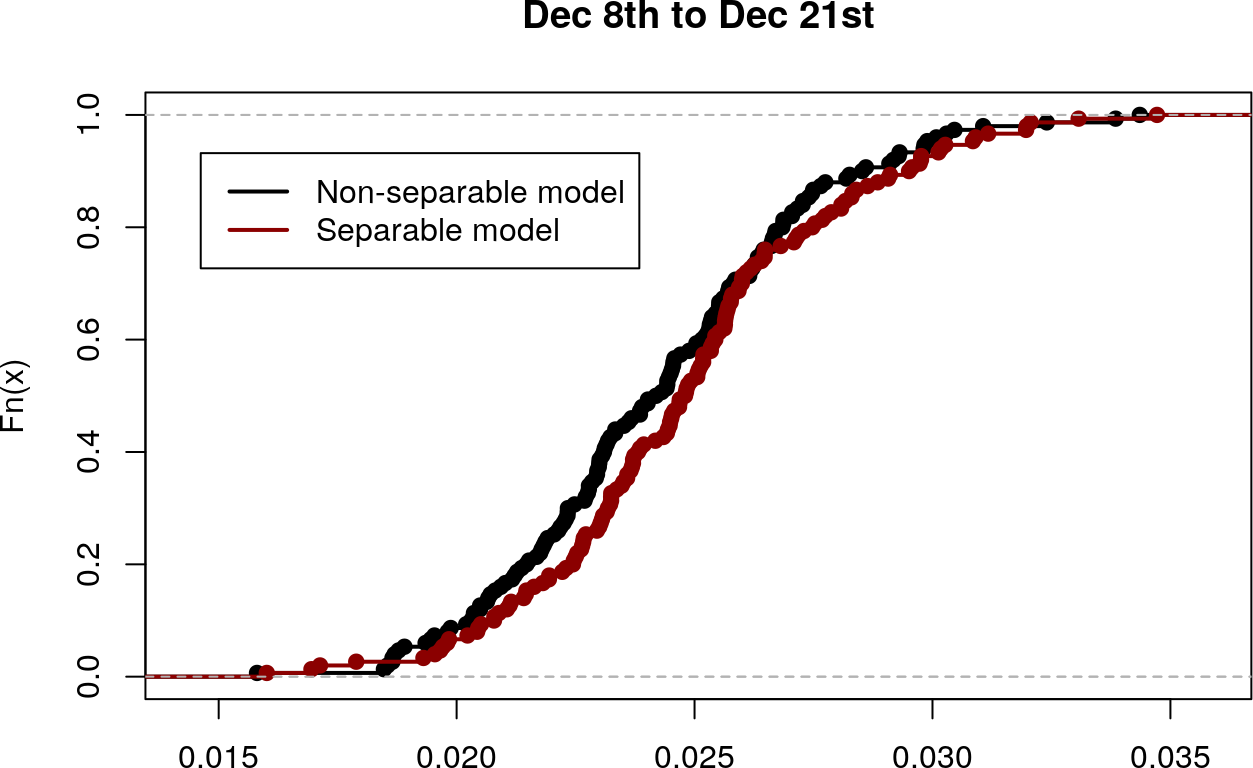}}
	\caption{More examples of comparisons between separable and non-separable approaches using ISE criterion.}
	\label{fig:ISE-5d-14d-test-separability}
\end{figure}

As already mentioned, the procedures described in Sections~6.1.1 and~6.1.2 were run considering additional time windows of 5 and 14 days respectively, and the results are presented hereinafter. Figure~\ref{fig:ISE-5d-14d-test-separability} displays a series of comparisons between the separable and non-separable approaches in terms of ISE criterion. We can observe in all situations the same behaviour as in Figure~15, which highlights the stability of our finding. There is only one time window when the two models perform equally, but we didn't find any scenario where the separable model outperforms the non-separable one. Table~\ref{tab:rISE-planar-network-additional} summarises the estimates of the rISE criterion observed when comparing the network and planar approaches. Similarly to the examples presented in the paper, we can notice that the average rISE for the network approach is several times smaller than the planar counterpart. 

\begin{table}
	\centering
	\caption{Additional numerical summaries of the comparisons between network and planar approaches using the rISE criterion defined in Equations~(12) and~(13) considering 6 time windows of 5 and 14 days.}
	\label{tab:rISE-planar-network-additional}
	\begin{tabular}{lc*{4}{S[
				round-mode = figures,
				scientific-notation = true
				]}}
		\toprule Time window & Type & {Mean} & {Std. Dev.} & {0.25 Quantile} & {0.75 Quantile}\\
		\midrule
		\multicolumn{6}{c}{5 days} \\
		\cmidrule{1-6}
		\multirow{2}{*}{Oct. 15th to Oct. 19th} 
		& Network & 446271.2 & 461537.9 & 179613.3 & 500750 \\
		& Planar & 78106937.8 & 72618775.1 & 32512752 & 101481168 \\
		\multirow{2}{*}{Nov. 15th to Nov. 19th} 
		c     & Network & 787433.4 & 1007141 & 229220 & 837075 \\
		& Planar & 77506206.8 & 128045868 & 23189588 & 90660632 \\
		\multirow{2}{*}{Dec. 15th to Dec. 19th} 
		& Network & 149465 & 183834 & 82689 & 134322 \\
		& Planar & 15766359 & 6347169.7 & 11390305 & 18086104 \\
		\cmidrule{1-6} \multicolumn{6}{c}{14 days} \\
		\cmidrule{1-6} \multirow{2}{*}{Oct. 9th to Oct. 22nd} 
		& Network & 14474853 & 19477536 & 1024963.1 & 10965064.4 \\
		& Planar & 969385104 & 1533389322 & 191835826 & 953339489\\
		\multirow{2}{*}{Nov. 9th to Nov. 22nd} 
		& Network & 596465.5 & 1758676 & 65439.33 & 476615.35 \\
		& Planar & 23773862.6 & 18583282 & 12362034 & 28534978 \\
		\multirow{2}{*}{Dec. 9th to Dec. 22nd} 
		& Network & 73423.16 & 17703.38 & 63242.50 & 80219.12 \\
		& Planar & 15904467.46 & 6315710.03 & 11939812 & 18567889 \\ 
		\bottomrule
	\end{tabular}
\end{table}

\end{refsection}
\newrefcontext[sorting=nty]
\printbibliography[section=2]
\end{document}